# A skeletal mechanism for premixed flames of isooctane/2-methylfuran blends

Atmadeep Bhattacharya[*]

Aalto University, School of Engineering, Department of Mechanical Engineering, 00076 Aalto, Finland

**Abstract**

The prospects of 2-methylfuran (2MF) as a bio-derived fuel that can be blended with gasoline are quite high. However, the effects of blending 2MF on practical combustion parameters like the laminar burning velocity and pollutant emission from gasoline/air mixture need to be assessed properly before the successful application of the bi-component fuel blend in spark ignition engines. Therefore, a skeletal chemical kinetic mechanism—containing 252 species and 1288 reactions—for the simulation of premixed flames involving isooctane (representing gasoline)/2MF blends has been proposed in the present work. The proposed model has been validated against a wide range of experimental data on the ignition delay time, laminar burning velocity and species profiles from burner stabilized flat premixed flames and well stirred reactors. The results obtained in the present work suggest that there are little chances of co-oxidation reactions among the species generated from the initial decomposition of both isooctane and 2MF. Furthermore, there is only around 8.6% variation in the peak laminar burning velocity when the mole fraction of 2MF is increased from 10% to 90% in isooctane/2MF blend. This fact is quite beneficial as this would entail minimum hardware modifications in the present SI engines that are designed for gasoline. It has also been observed in the present study that the presence of 2MF does not influence the peak benzene mole fraction in the isooctane/air flame at rich conditions. However, the NO mole fraction increases with the increase in 2MF quantity in isooctane/2MF blend.

[*] Corresponding author. E-mail address: atmadeep.bhattacharya@aalto.fi

# 1. Introduction

The contemporary transportation sector accounts for around 20% of the global fuel consumption and is primarily driven by liquid petroleum oils [1]. The use of bio-derived liquid fuels is a promising strategy towards the abatement of the share of fossil fuels in this sector. In this regard, the second-generation biofuels such as 2-methylfuran and 2,5-dimethylfuran are considered as propitious alternatives to gasoline in spark ignition (SI) engines [2,3]. These furanic unsaturated cyclic ethers can be produced from lignocellulosic bio-wastes and therefore do not threaten the food security [4,5]. The storage and handling issues of these fuels are also simplified by the fact that they are negligibly soluble in water. However, the emissions of harmful polycyclic aromatic hydrocarbons are far higher in 2,5-dimethylfuran flames compared to the ones involving 2-methylfuran (2MF) [6]. Furthermore, the advantages of 2MF as a biofuel include high research octane number (100.7), high energy content (30.37 MJ/kg), low boiling point (64°C) and low enthalpy of vaporization (358 kJ/kg) [7]. These properties make 2MF a member of the ''Tailor-Made Fuels from Biomass" family [5,8].

Many a researcher has exploited these favourable properties of 2MF in spark ignition (SI) engines. Thewes et al. [7] experimentally measured the engine performance of 2MF in a direct injection SI (DISI) engine. The results were compared with the performance obtained using a gasoline surrogate fuel and ethanol which is the most common renewable liquid fuel for SI engines. It was observed in this study that 2MF provided stable combustion during the cold start of the engine due to its low boiling point. Therefore, the hydrocarbon emission reduced greatly (~61%) compared to the gasoline fuel at similar operating conditions. These observations have been corroborated by Hoppe et al. [9] in a more recent study. Furthermore, it was observed that 2MF considerably increased the knock resistance in a DISI engine. In another work, Wang et al. [10] observed that 2MF has the ability to improve the indicated thermal efficiency of a DISI engine by about 3% compared to conventional gasoline. Moreover, most of the emissions were found to be reduced when 2MF was used in the engine. It may be mentioned in this regard that a blend of 10% 2MF and 90% gasoline has recently been tested in a successful 90,000 km road trial in a conventional SI engine [2] without any significant damage to the engine. Based on these observations, it has been presumed in this work that 2MF and gasoline blends can safely be used in the conventional SI engines without further design modifications.

However, some important fundamental combustion parameters like the ignition delay time and laminar burning velocity of the fuel blends in air need to be prudently evaluated and analysed at engine relevant conditions a priori for efficient operation of SI engine using 2MF /gasoline blends. In this regard, there has been quite a few works on the exclusive experimental determination of ignition delay time and laminar burning velocity of both isooctane [11–14] and 2MF [15–17] at different physical conditions. However, the complexities associated with different experimental methods mandate numerical determination of these fundamental parameters using chemical kinetic mechanisms. The contemporary review works done by Sarathy et al. [18] and Zhen et al. [19] have shown that there is a plethora of chemical kinetic mechanisms of gasoline surrogates in the literature that have been validated across a wide range experimental observations from apparatus like shock tube, rapid compression machine, different types of burners (flat flame, counterflow, Bunsen etc.), different types of reactors (perfectly stirred, plug flow etc.), combustion bomb, homogeneous charge compression ignition (HCCI) engines etc. As mentioned in these works, the binary mixtures of isooctane and n-heptane are quite commonly used as the gasoline surrogates. In this regard, Bhattacharya et al. [20] considered a binary mixture of 95% isooctane and 5% n-heptane as a surrogate for the real world gasoline. Furthermore, Huang et al. [21] showed that the laminar burning velocity of the isooctane/n-heptane blend does not vary significantly when the isooctane quantity is varied in the range of 85-100 %. Therefore, the exploration and proposal of new isooctane combustion chemistry still remains an extremely important topic [22–24].

On the other hand, Simmie and Curran [25] performed quantum chemical calculations to determine the enthalpies of formation and bond dissociation energies of alkylfuran compounds. Somers et al. [26] proposed a detailed chemical kinetic mechanism for 2MF oxidation containing 2059 reactions among 391 species. The mechanism was validated using the experimental data on ignition delay time and laminar burning velocity measured using shock tube and heat flux burner [27] respectively at various equivalence ratios ($\varphi$) in the high temperature regime. The pyrolysis chemistry of this mechanism was later updated using quantum chemical methods [28]. In recent times, Cheng et al. [29] have proposed a revised version of this mechanism containing the formation and decomposition pathways of propargyl, fulvene, benzene, benzyl, styrene, ethylbenzene etc. Furthermore, a group of researchers at Bielefeld

University and Université de Lorraine [30–32] have developed another detailed kinetic model consisting of 1472 reactions among 305 species for furanic fuels—furan, 2,5-dimethylfuran and 2MF—based on the mechanism proposed by Sirjean et al. [33]. This high temperature model was validated against the experimental data on the intermediate species profiles obtained using the high-resolution electron–ionization molecular beam mass spectrometry (EI-MBMS) and gas chromatography (GC) of laminar premixed flat flames involving 2MF, oxygen and argon at stoichiometric as well as fuel rich conditions at low pressures (20 and 40 mbar). In a subsequent work, this mechanism was augmented with polycyclic aromatic hydrocarbons (PAH) and species often considered in gasoline surrogate mixtures (i.e. toluene, n-heptane and isooctane) [6]. Furthermore, this mechanism was improved with low to moderate temperature combustion chemistry recently which finally resulted in a size of 3143 reactions among 524 species [34]. In another recent work, Weber et al. [35] studied the pyrolysis of furan, 2MF and 2,5-dimethylfuran in shock tube. The experimentally obtained temporal variations of H atom concentration were simulated using a modified version of the mechanism by Liu et al. [30]. In this modified version, the specific reaction rates of 17 influential reactions steps were altered.

However, during the computational fluid dynamics (CFD) simulation of reactive flows, it is the associated chemical mechanism that demands the lion's share of the computational resources. In this regard, Lu et al. [36] have expressed the overall cost of a CFD simulation ($C_o$) as, $C_o = \sum_{n=1}^{4} \Delta_n N^{n-1}$ where, "$N$" is the number of species in the mechanism and $\Delta$'s are the case specific constants that depend on the determination of parameters like reaction rate, mass diffusion, Jacobian factorization and number of nodes in the computational domain. Therefore, in order to keep the computational cost within the achievable limit, numerous approaches have been adopted to reduce the number of species in a CFD simulation [37,38]. Some of the popularly accepted techniques include the development of skeletal and reduced chemical mechanisms [39,40], invariant constrained equilibrium edge preimage curve method (ICE-PIC) [41], in situ adaptive tabulation (ISAT) [42], flamelet generated manifold (FGM) [43], rate-controlled constrained equilibrium (RCCE) [44] and many more.

As seen from the above literature review, even though isooctane/2MF blend has great potential in the automobile sector, few works are available on the effects of adding 2MF with isooctane from the perspective of chemical kinetics and pollutant formation in premixed flames. Moreover, concise mechanisms are needed for such studies in order to keep the computational costs within acceptable limits. Therefore, a skeletal chemical kinetic mechanism has been developed in the present work for such a purpose. This mechanism has been validated using the ignition delay time, laminar burning velocity and species profiles from experimentally investigated premixed flames involving isooctane, 2MF and their blends from literature in the present work. Furthermore, the proposed mechanism has been used to assess the effects of 2MF blending to isooctane from the soot precursor and NO emission point of view.

## 2. Kinetic model development

The starting point for the development of the skeletal mechanism has been chosen to be the detailed mechanism developed at the National University of Ireland Galway [26,28,45]. This mechanism (hereafter called NUIG mechanism) contains 2889 reactions among 567 species. This mechanism has been validated against a wide range of experimental data on ignition delay time from shock tube, laminar burning velocity from heat flux burner and species profiles from jet stirred reactor. However, in order to achieve a concise version of the mechanism, the detailed NUIG mechanism has been reduced to a skeletal mechanism with 590 reactions among 113 species using the directed relation graph with error propagation (DRGEP) method [46]. During the reduction process, the target parameters have been chosen to be the high temperature ignition delay times of $2MF/O_2/Ar$ mixtures at different pressures (1.25, 4.25 and 10.65 bar) and equivalence ratios ($\varphi$ = 0.25, 1 and 2) from the work of Wei et al. [16]. A thorough description of a similar reduction process can be found in Bhattacharya et al. [47].

After obtaining the skeletal mechanism for 2MF oxidation, another mechanism for the combustion of isooctane from Yoo et al. [48] has been merged into the former in order to construct a kinetic model for isooctane/2MF blend. In this context, it may be mentioned that all the chemical kinetic parameters in these mechanisms are expressed in Arrhenius form. During the merger process, the chemical kinetic parameters belonging to the isooctane mechanism have been given preference over the 2MF mechanism in case of conflicts. On the other hand, the thermodynamic (polynomial fits to the specific heats, standard state enthalpies, and standard state entropies) and transport parameters (the Lennard-Jones parameters, dipole moments, polarizabilities and rotational relaxation collision numbers) belonging to the 2MF mechanism have been preferred over the same for the isooctane mechanism. The primary motivation behind these choices has been the betterment of the validation results. On the other hand, Alexandrino et al. [49] have concluded in their work that the nitric oxide (NO) chemistry has significant influences on the 2MF combustion kinetics over a wide range of equivalence ratios. Therefore, this NO mechanism [49] —both the thermal and prompt routes—has been added to the proposed chemical kinetic model. At the end of this step, the size of the mechanism obtained was 1293 reactions among 252 species including the inert gases (Argon and Helium).

However, due to the replacement of a large number of reactions in the 2MF mechanism with the ones from the isooctane mechanism, considerable over prediction has been observed in predicting the laminar burning velocity of 2MF/air mixtures with this mechanism. The experimental data set from Somers et al. [26] has been used as target parameters in this context. In order to rectify such mismatch, the chemical kinetic parameters of the proposed skeletal mechanism have been further optimized using sensitivity analysis on the mass burning flux for laminar premixed flames. During the sensitivity analysis, a normalized sensitivity coefficient for a reaction '$i$' with specific forward reaction rate $k_i$ is defined as $\partial \ln \dot{m} / \partial \ln \dot{k_i}$, where $\dot{m}$ represents the mass burning flux. It may be noted in this regard that the mass burning flux is directly proportional to the laminar burning velocity for a fuel/air mixture at constant pressure and unburnt gas temperature. Based on the results from the sensitivity analysis, the specific reaction rate parameters pertaining to four reactions have been altered. The details of these changes are listed in Table 1. Furthermore, five reactions (viz. $CH_3O_2 + CH_3 = CH_3O + CH_3O$, $CH_3O_2 + H = CH_3O + OH$, $CH_3O_2 + O = CH_3O + O_2$, $CH_3 + O_2 (+M) = CH_3O_2 (+M)$ and $HCCO + O_2 = CO_2 + CO + H$) have been excluded from the mechanism that was formed earlier. Therefore, the final version of the proposed mechanism consists of 1288 reactions among 252 species.

## 3. Validation of the mechanism

After the formulation of the mechanism, the next step has been to validate it against the experimental data from literature. The high temperature ignition delay time dataset for 2MF/$O_2$/Ar blends from Wei et al. [16] has been used for that purpose initially since the same had been used as the target during the mechanism reduction process. Furthermore, the performance of the proposed mechanism in predicting the ignition delay time ($\tau$) has been compared with the detailed NUIG mechanism [26,28,45]. The results are shown in Fig. 1 (a-g) for different compositions ($\varphi = 0.25$, 1 and 2) and pressures (viz. 1.25 bar, 4.25 bar and 10.65 bar). It may be seen from the figure that despite containing around 55% less number of species compared to the detailed one, the proposed mechanism predicts the experimental results accurately. On a similar note, the laminar burning velocities ($S_L$) predicted by the present mechanism for 2MF/air mixtures at atmospheric pressure and different

Table 1. Changes in the Arrhenius parameters applied to the present mechanism (Unit system: cm3, mol, s, cal).

| Reaction | NUIG mechanism | | | Present mechanism | | | Ref. |
|---|---|---|---|---|---|---|---|
| | A | n | $E_a$ | $A_{new}$ | $n_{new}$ | $E_{a\_new}$ | |
| $H+O_2(+M) = HO_2(+M)$ | High pressure limit | | | | | | [50] |
| | $1.475\times10^{12}$ | 0.6 | 0 | $5.58\times10^{12}$ | 0.4 | 0 | |
| | Low pressure limit | | | | | | |
| | $3.482\times10^{16}$ | -0.411 | -1115 | $84\times10^{16}$ | -0.8 | 0 | |
| | TROE parameters | | | | | | |
| | $\alpha=0.5$, $T^{***}=1.0\times10^{-30}$, $T^*=1.0\times10^{30}$, $T^{**}=1.0\times10^{10}$ | | | $\alpha=0.5$, $T^{***}=1.0\times10^{-30}$, $T^*=1.0\times10^{30}$, $T^{**}=0$ | | | |
| | Third body efficiency | | | | | | |
| | $\Omega_{CH4}=2.0$, $\Omega_{C2H6}=3.0$, $\Omega_{CO}=1.9$, $\Omega_{CO2}=3.8$, $\Omega_{H2}=1.3$, $\Omega_{H2O}=14.0$ | | | $\Omega_{Ar}=0.8$, $\Omega_{CO}=1.2$, $\Omega_{CO2}=2.4$, $\Omega_{O2}=0.0$, $\Omega_{H2}=2.5$, $\Omega_{H2O}=18.00$, $\Omega_{N2}=1.26$ | | | |
| $CH_3+H(+M) = CH_4(+M)$ | High pressure limit | | | | | | [51] |
| | $1.27\times10^{16}$ | -0.6 | 383 | $1.20\times10^{15}$ | -0.4 | 0 | |
| | Low pressure limit | | | | | | |
| | $2.48\times10^{33}$ | -4.76 | 244 | $6.40\times10^{23}$ | -1.8 | 0 | |
| | TROE parameters | | | SRI parameters | | | |
| | $\alpha=7.83\times10^{-1}$, $T^{***}=74.0$, $T^*=2940.0$, $T^{**}=6960.0$ | | | a=0.45, b=797.0, c=979.0 | | | |
| | Third body efficiency | | | | | | |
| | $\Omega_{CH4}=2.0$, $\Omega_{C2H6}=3.0$, $\Omega_{CO}=1.5$, $\Omega_{CO2}=2.0$, $\Omega_{H2}=2.0$, $\Omega_{H2O}=6.0$ | | | $\Omega_{CO}=2.0$, $\Omega_{CO2}=3.0$, $\Omega_{H2}=2.0$, $\Omega_{H2O}=5.0$ | | | |

(Table continued)

| Reaction | NUIG mechanism | | | Present mechanism | | | Ref. |
|---|---|---|---|---|---|---|---|
| | A | n | $E_a$ | $A_{new}$ | $n_{new}$ | $E_{a\_new}$ | |
| HCO+M=H+CO+M | $4.75\times10^{11}$ | 0.66 | $1.49\times10^4$ | $1.2\times10^{17}$ | -1 | $17\times10^3$ | [51] |
| | Reverse Reaction | | | | | | |
| | $3.582\times10^{10}$ | 1.041 | -457.3 | $\Omega_{CH4}$=2.80, $\Omega_{H2O}$=5.00, $\Omega_{CO2}$=3.00, $\Omega_{H2}$=1.90, $\Omega_{CO}$=1.90 | | | |
| | $\Omega_{H2}$=2.0, $\Omega_{H2O}$=12.0, $\Omega_{CO}$=1.5, $\Omega_{CO2}$=2.0, $\Omega_{CH4}$=2.0, $\Omega_{C2H6}$=3.0 | | | | | | |
| MF2=MF22J+H | $3.37\times10^{15}$ | -0.01 | 87487.9 | $2.5260\times10^{12}$ | 0.898 | $8.5399\times10^4$ | [52] |
| | Pressure Dependence Through Logarithmic Interpolation (PLOG) | | | | | | |
| | Pressure = 0.01 atm | | | Pressure = 0.001 atm | | | |
| | $2.886\times10^{92}$ | -22.81 | $1.27594\times10^5$ | $2.62\times10^{62}$ | -14.96 | 100667 | |
| | Pressure = 0.1 atm | | | Pressure = 1 atm | | | |
| | $1.064\times10^{87}$ | -20.85 | $1.28912\times10^5$ | $1.37\times10^{60}$ | -13.03 | 112531 | |
| | Pressure = 1.0 atm | | | Pressure = 2 atm | | | |
| | $1.416\times10^{65}$ | -14.34 | $1.1791\times10^5$ | $1.44\times10^{54}$ | -11.26 | 109567 | |
| | Pressure = 2.5 atm | | | Pressure = 5 atm | | | |
| | $2.149\times10^{55}$ | -11.48 | $1.12302\times10^5$ | $3.12\times10^{45}$ | -8.71 | 104865 | |
| | Pressure = 10 atm | | | Pressure = 10 atm | | | |
| | $1.629\times10^{41}$ | -7.39 | $1.03833\times10^5$ | $7.09\times10^{38}$ | -6.77 | 101049 | |
| | Pressure = 100 atm | | | Pressure = 20 atm | | | |
| | $1.97\times10^{24}$ | -2.53 | $9.32166\times10^4$ | $3.50\times10^{32}$ | -4.94 | 97304 | |
| | | | | Pressure = 50 atm | | | |
| | | | | $1.75\times10^{25}$ | -2.84 | 92847 | |
| | | | | Pressure = 100 atm | | | |
| | | | | $5.76\times10^{20}$ | -1.55 | 90053 | |
| | | | | Pressure = 1000 atm | | | |
| | | | | $3.24\times10^{12}$ | 0.82 | 84802 | |

temperatures (298 K, 328 K, 358 K, 398 K) for $\varphi$ = 0.55-1.65 also agree excellently with the experimental observations—made using the heat flux burner—from Somers et al. [26] in Fig. 2. However, it may be seen from the figure that the maximum value of $S_L$ occurs at $\varphi$ = 1.15 for the simulation results while the corresponding experimental value is $\varphi$ = 1.1.

Similar to the heat flux burner, the McKenna burner also produces laminar flat one-dimensional premixed flames. The major species profiles from such flat flames of 2MF/O$_2$/Ar mixtures as obtained by Cheng et al. [29] and Tran et al. [31] have been used as the next validation targets. During this process, the experimentally measured temperature profiles along the height above the burner (HAB) for three different flames (with $\varphi$ = 0.8, 1.0 and 1.7) have been provided as inputs and only the species conservation equations have been solved using the proposed chemical kinetic mechanism. The variation of the mole fractions of seven major species is plotted along the HAB for $\varphi$ = 0.8 in Fig. 3 (a). It may be seen from the figure that the predictions by the proposed chemical kinetic mechanism are within the uncertainty limits of the experimental values from Cheng et al. [29]. However, as observed from the figure, the experimental values for 2MF and $O_2$ have not been evaluated accurately at the burner surface. Therefore, the experimental data for six major species have been taken from Tran et al. [31] for $\varphi$ = 1.0 and 1.7 respectively in Fig. 3 (b and c). As seen from these figures, the model predictions match excellently with the experimental results. It may be noted in this regard that the comparison of the proposed model performance in predicting major species profiles in the flame with the experimental data from Cheng et al. [29] at stoichiometric condition and $\varphi = 1.5$ has been placed in the supplementary material (Fig. S1 (a) and (b)). Additionally, some of the important intermediate species profiles (viz. methyl radical ($CH_3$), methane ($CH_4$), formaldehyde ($CH_2O$), acetylene ($C_2H_2$), ethylene ($C_2H_4$), ethyl radical ($C_2H_5$), ethane ($C_2H_6$), sum of allene and propyne ($C_3H_4$), allyl radical ($C_3H_5$), propene ($C_3H_6$), 1,3-butadiyne ($C_4H_2$), 1-butene-3-yne ($C_4H_4$), sum of 1,3-butadiene, 1,2-butadiene, and 2-butyne ($C_4H_6$) and benzene ($C_6H_6$)) along the height above the burner plate have been validated against the experimental data from Tran et al. [31] in Fig. S2 and S3 (a-n) for $\varphi = 1.0$ and 1.7 respectively. It may be seen from the figures that the match is qualitative for most of the species. A better prediction would entail a more detailed mechanism in lieu of the skeletal one used in the present work.

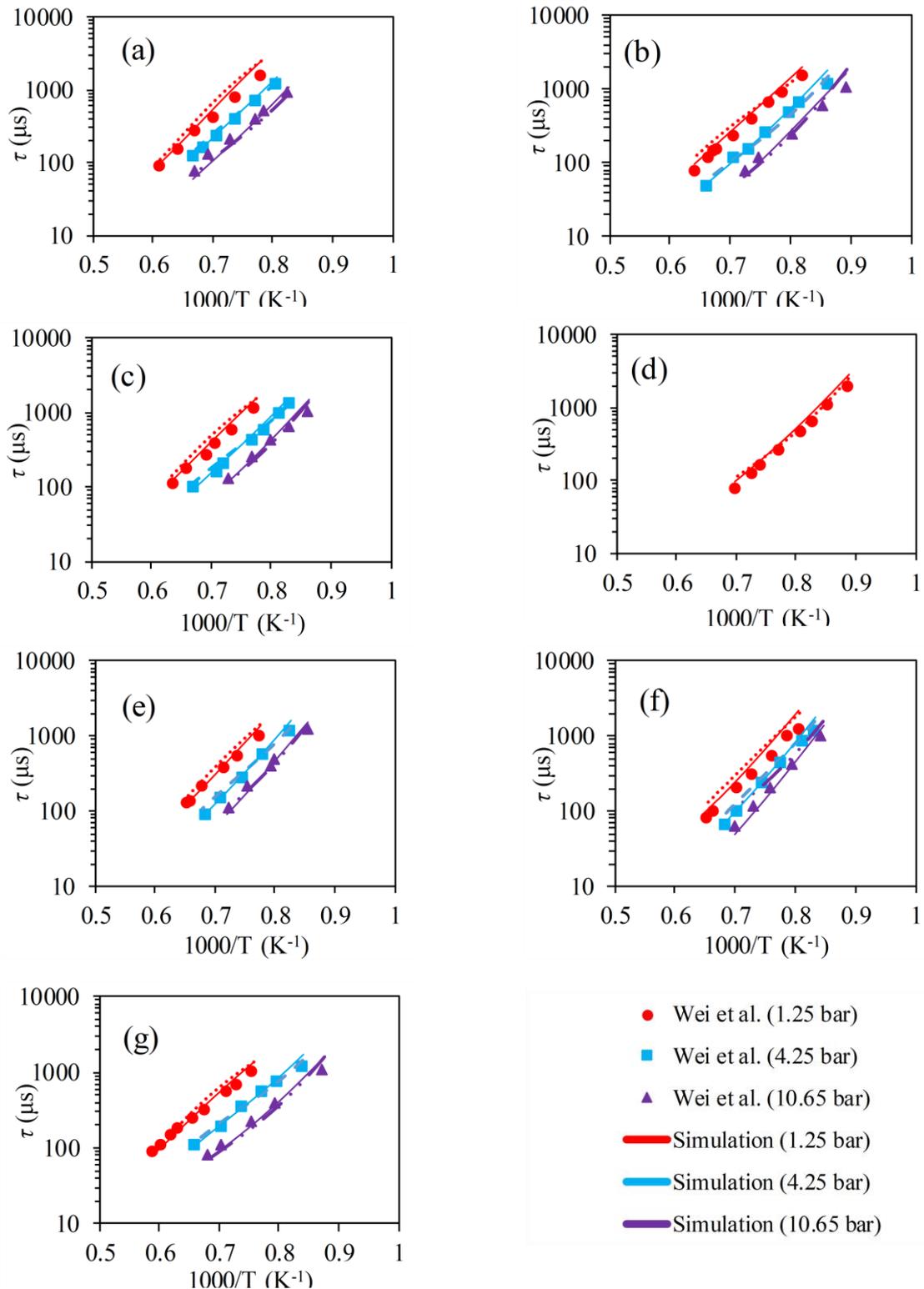

Figure 1. Comparison of performance of the proposed mechanism (dotted lines) against the detailed NUIG mechanism (bold lines) [26,28,45] and experimental data (symbols) from Wei et al. [16] towards the prediction of high temperature ignition delay times for compositions (2MF: $O_2$: Ar) (a) 752: 4511: 94737 ($\varphi$ = 1), (b) 753: 4518: 44729 ($\varphi$ = 1), (c) 1: 6: 93 ($\varphi$ = 1), (d) 3383:20298:76319 ($\varphi$ = 1), (e) 503: 6030: 93467 ($\varphi$ = 0.5), (f) 252: 6045: 93703 ($\varphi$ = 0.25), (g) 1980: 5941: 92079 ($\varphi$ = 2) at different pressures (1.25, 4.25 and 10.65 bar).

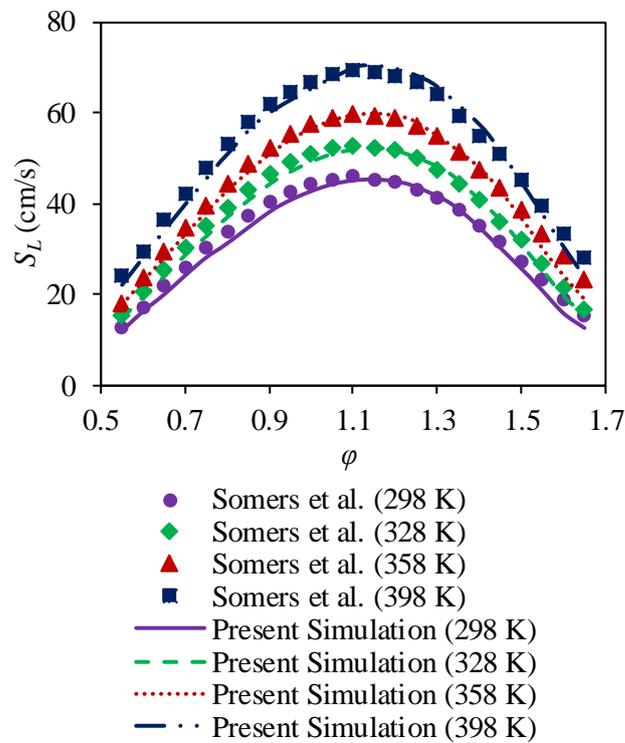

Figure 2. Laminar burning velocities of 2MF/air mixtures at atmospheric pressure and different temperatures (298 K, 328 K, 358 K, 398 K) for $\varphi$ = 0.55-1.65. The experimental data are taken from Somers et al. [26].

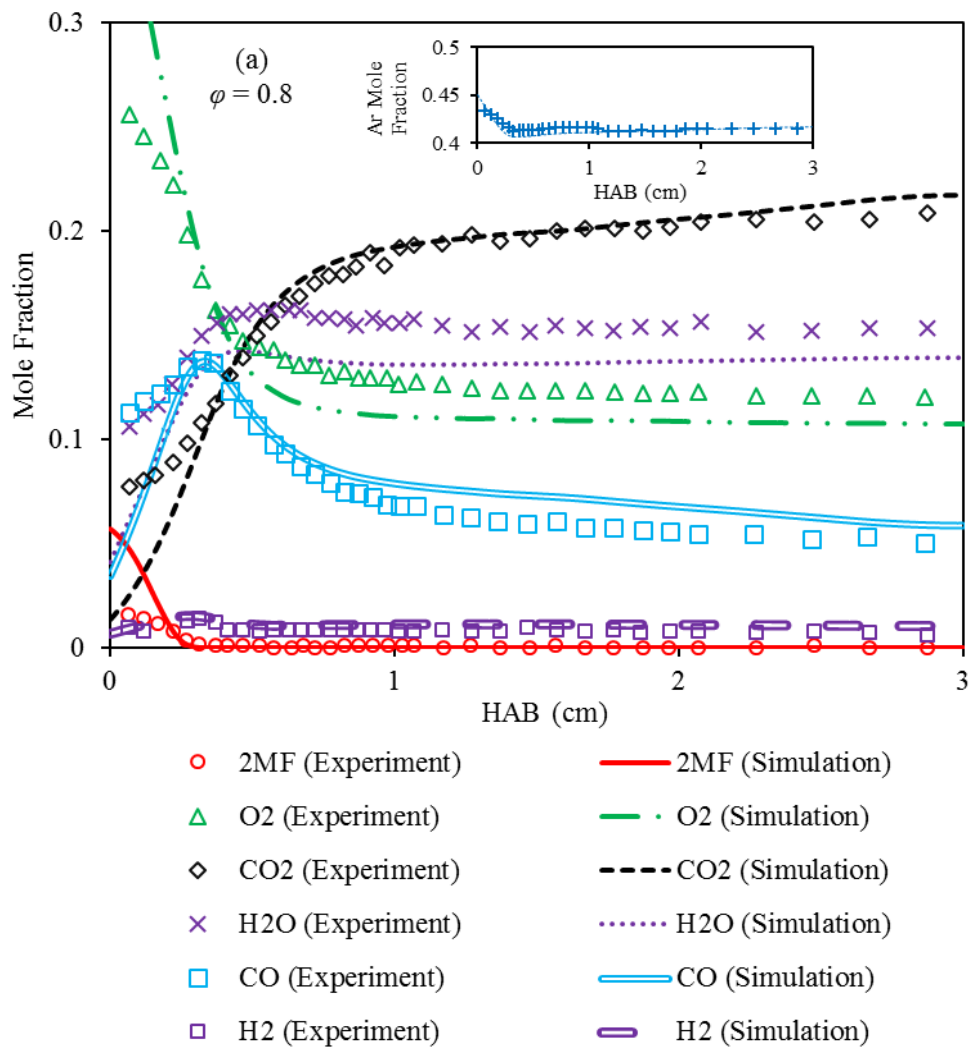

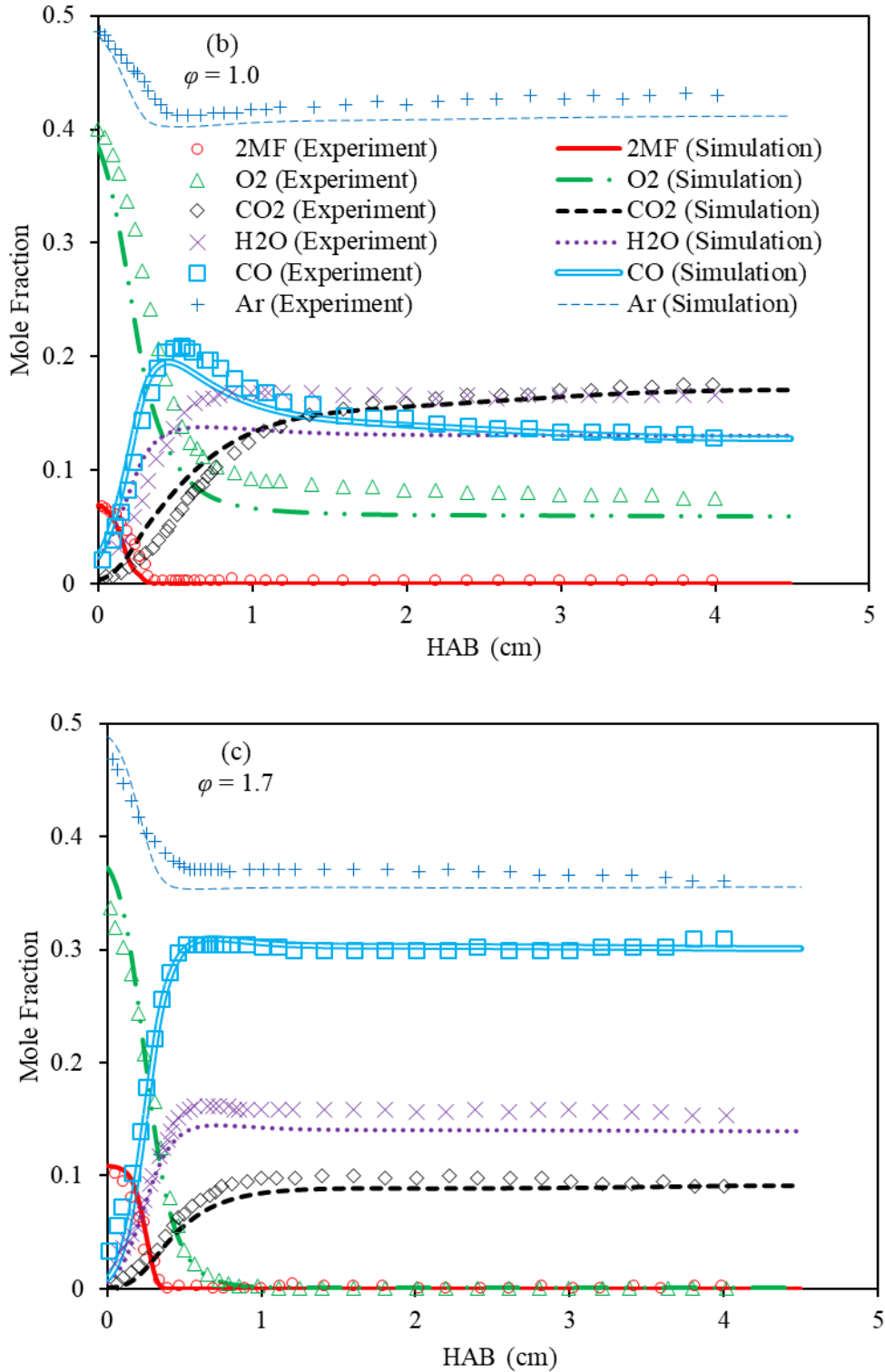

Figure 3. Comparison of model prediction of major species profiles for flat 2MF/air flames stabilized above McKenna burner for (a) $\varphi = 0.8$ with experimental data from Cheng et al. [29], (b) $\varphi = 1.0$ with experimental data from Tran et al. and (c) $\varphi = 1.7$ with experimental data from Tran et al. [31]. The legends of (b) and (c) are same.

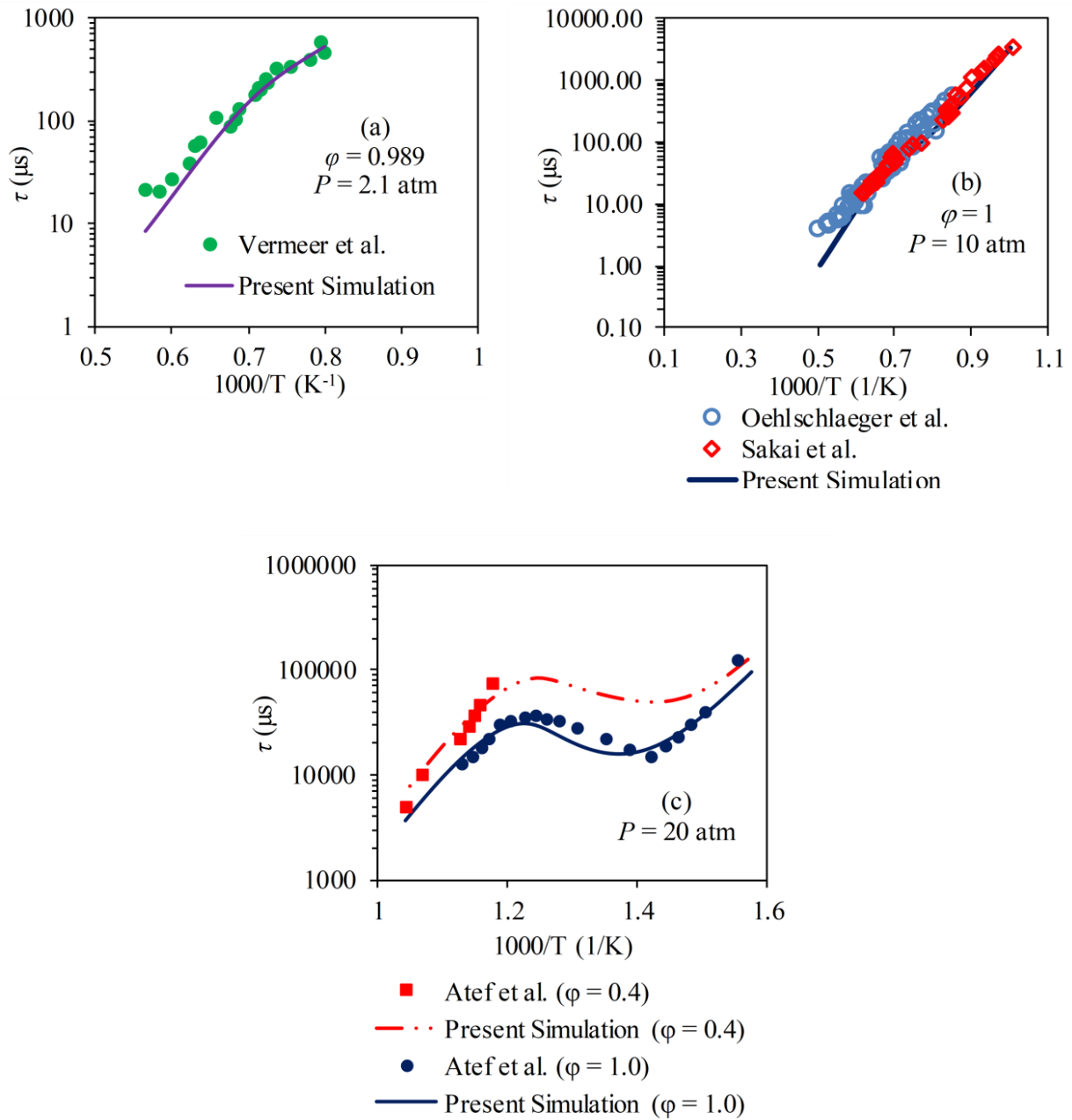

Figure 4. Comparison between the predicted ignition delay time using the proposed mechanism at different temperatures and equivalence ratios with the experimental values [53–56] at (a) 2.1 atm pressure and isooctane/$O_2$/Ar mixture, (b) 10 atm pressure and isooctane/air mixture, (c) 20 atm pressure and isooctane/$O_2$/$N_2$ mixture.

Similarly, qualitative agreement has been observed for the model predictions and the experimental data [57] for isooctane/$O_2$/$N_2$ mixtures from jet stirred reactor at $\varphi$ = 0.5, 1 and 1.5 in the temperature range 700-1100 K, 1 s residence time and 10 atm pressure. The results may be seen in Fig. S4 (a-c). On the other hand, Fig. 4 (a-c) compares the predicted ignition delay time with the experimental data for a wide range of temperature, pressure and equivalence ratio. In Fig. 4 (a), the simulated ignition delay time at pressure (*P*) 2.1 atm and temperature (*T*) range 1250-1770 K have been compared with the experimental data from Vermeer et al. [54] for near stoichiometric ($\varphi$ = 0.989) isooctane/$O_2$/Ar mixture. Moreover, the simulated ignition delay time at *P* = 10 atm and *T* in the range 1000-1980 K have been compared with the experimental data from Oehlschlaeger et al. [53] and Sakai et al. [55] for stoichiometric isooctane/air mixture in Fig. 4 (b). These two figures confirm that the present model predicts the high temperature ignition delay time of isooctane at stoichiometric condition quite accurately. The ignition delay time at $\varphi$ = 0.4 and low temperature range is validated in Fig. 4 (c) against the experimental data—obtained using the rapid compression machine (RCM) at University of Connecticut— from Atef et al. [56]. It may be seen from the figure that along with predicting the ignition delay time at lean premixed condition ($\varphi$ = 0.4), the proposed mechanism performs excellently in predicting the negative temperature coefficient (NTC) [58] region as well. One of the salient characteristics of the NTC flame chemistry in the intermediate temperature range (≈ 750-850 K at high pressure) is the increased importance of the reaction $C_8H_{17}+O_2 = C_8H_{16}+HO_2$. The comparatively stable hydroperoxyl radicals ($HO_2$)—compared to the highly reactive H, OH and O radicals— contribute towards the decrease in the reactivity of the gas mixture [47,59].

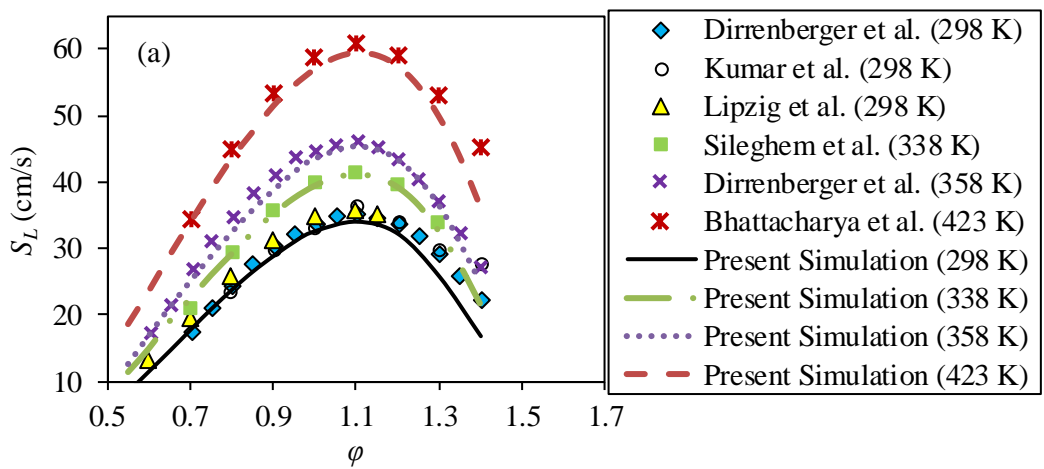

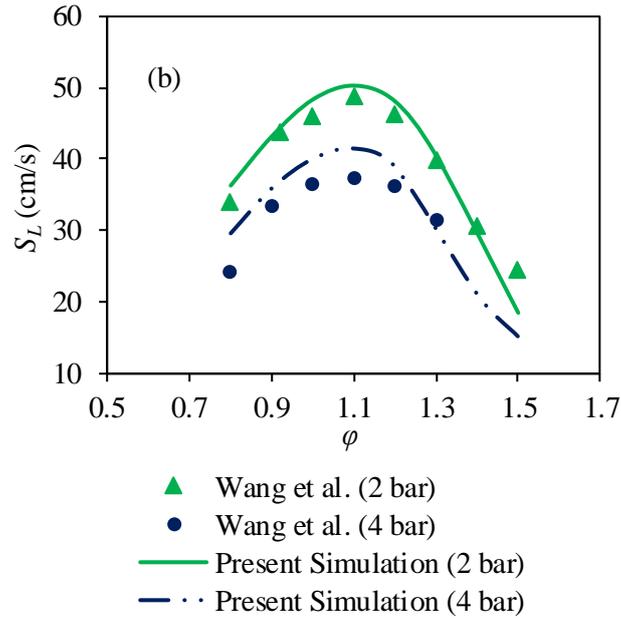

Figure 5. Assessment of performance of the proposed mechanism in predicting the variation of laminar burning velocity of isooctane/air mixture at different equivalence ratios ($\varphi$), unburnt gas temperatures and (a) atmospheric pressure, (b) 2 bar and 4 bar pressure. The experimental data at atmospheric pressure are taken from [60–64] and higher pressures are taken from [65].

For further validation of the proposed kinetic model, the experimental data from literature on the laminar burning velocity of isooctane/air mixture at various equivalence ratios, unburnt gas temperatures and pressures have been used. Among the experimental data, the atmospheric pressure values are obtained by the heat flux [20,61–63] and counter flow burner method [64] and the higher pressure values by outwardly propagating spherical flames in a combustion bomb [65]. It may be noted in this regard that different methods for the experimental determination of laminar burning velocity differ considerably in operating pressure range and accuracy [66]. Furthermore, the heat flux burner method is the most accurate one in the low pressure (≤10 bar) range [67]. It may be seen from Fig. 5 (a) that the proposed mechanism predicts $S_L$ of isooctane/air mixture at atmospheric pressure and different unburnt gas temperatures quite accurately in the lean side of stoichiometry. However, there is considerable under prediction in the rich side at temperatures 298 K and 423 K. There are differences in opinions about the actual cause of such mismatch between the model predictions and experimental data from heat flux burner for heavy alkanes like isooctane. Dirrenberger et al. [61] have mentioned that the increased importance of the ethylene and $C_4$ sub mechanisms is instrumental in creating such under predictions. Therefore, in their opinion, the corresponding kinetic parameters of these sub mechanisms need further revision for better performance of the kinetic models in the rich side. On the other hand, Egolfopoulos et al. [67] have pointed out that the experimental apparatus itself tends to give higher values of laminar burning velocity in the rich side compared to other methods. However, no reasons were mentioned for this discrepancy in this review work. In a subsequent study, Bhattacharya et al. [20] observed that the uncertainties originating from thermocouple placement aggravate at higher equivalence ratios beyond the stoichiometric point. It may be noted in this context that the rectification of the disparity in theoretical predictions and experimental data for rich isooctane/air laminar burning velocity is outside the scope of the present work.

Figure 5 (b) compares the model predictions for $S_L$ of isooctane/air mixture with the experimental data from Wang et al. [65] at higher pressures (i.e. 2 and 4 bar). It is noteworthy in this context that the accuracy of the combustion bomb has been under strong criticism in recent past [68,69]. Therefore, the achieved proximity between the experimental and numerical results in Fig. 5 (b) has been considered as successful validation of the proposed model in the present work. On a similar note, the model predictions of the $S_L$ compare satisfactorily with the corresponding experimental data of outwardly propagating spherical flames from Ma et al. [70] for isooctane/2MF (4:1 and 1:1) blends in air at atmospheric pressure, different unburnt gas temperatures and in the range $0.8 \leq \varphi \leq 1.4$ in Fig. 6 (a) and (b) respectively. These experimental values are the only set available in the literature for isooctane/2MF blend. In order to ascertain the uncertainty associated with the results obtained using outwardly propagating spherical flames in Fig. 6, the experimental values of $S_L$ of isooctane/air mixture from Ma et al. [70] are compared with the corresponding data set from Sileghem et al. [62] and Dirrenberger et al. [61] at similar physical conditions in Fig. S5. Sileghem et al. [62] and Dirrenberger et al. [61] adopted the heat flux burner method in their study. As seen from the figure, the values of Ma et al. [70] differ significantly from the other two sets of data. These deviations further corroborate the observation made by Chen [69] that the dataset on the laminar burning velocity from constant volume combustion chamber method should not be used as quantitative validation targets for the chemical mechanisms when $\varphi > 1.2$.

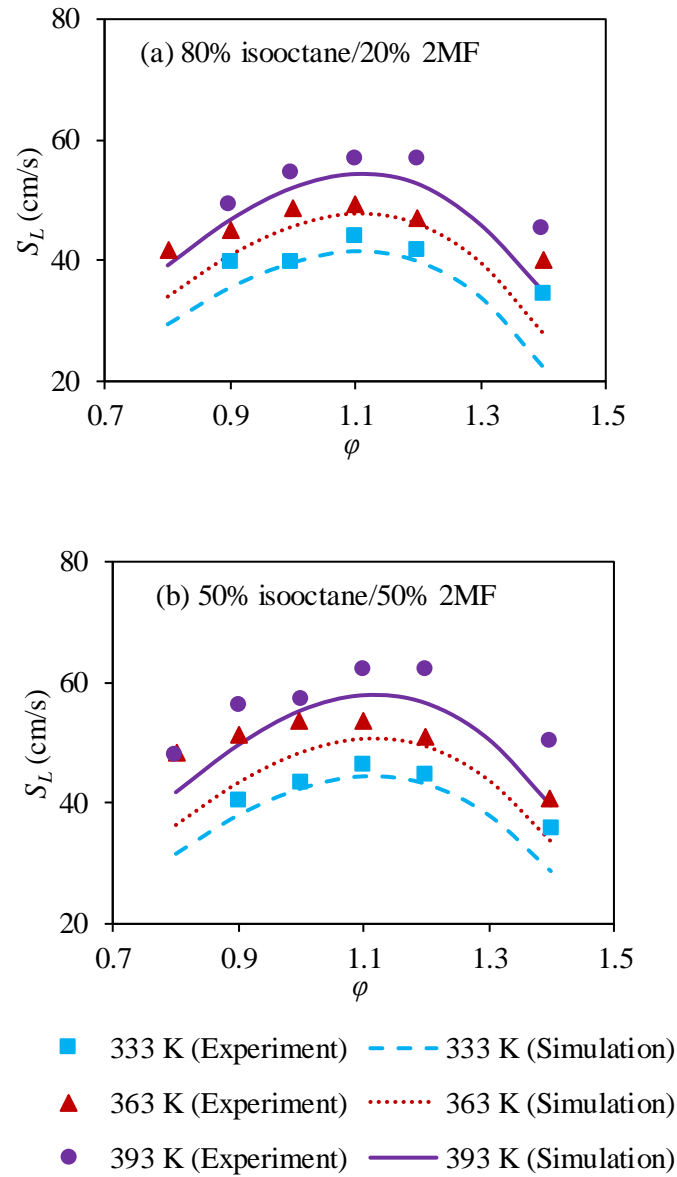

Figure 6. Assessment of performance of the proposed mechanism in predicting the variation of laminar burning velocity of (a) 80% isooctane/20% 2MF and (b) 50% isooctane/50% 2MF blends in air at different unburnt gas temperatures and equivalence ratios. The experimental data are taken from Ma et al. [70].

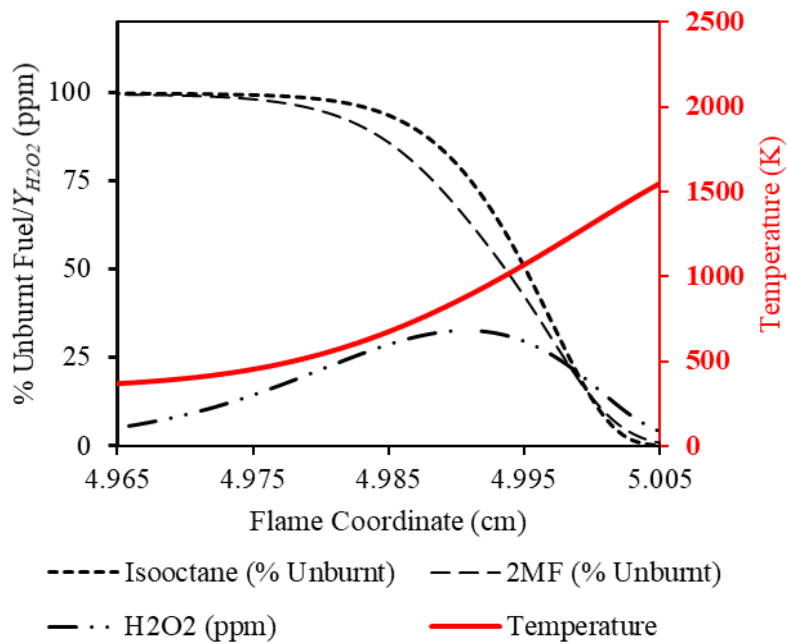

Figure 7. Variation of unburnt fractions of isooctane and 2MF along the flame coordinate for stoichiometric 80% isooctane/20% 2MF blend in air at 333 K unburnt gas temperature and 1 atm pressure. The variation of $H_2O_2$ mole fraction in ppm ($Y_{H2O2}$) in the same computational domain is also shown. The corresponding flame temperature variation is shown in the secondary axis (red).

## 4. Results and discussion

### 4.1 Analysis of the possibilities of co-oxidation reactions

After the validation of the proposed mechanism for the individual fuel components and their blends, the possibility of co-oxidation reactions has been explored. During the co-oxidation reactions, the initial heavy radicals generated from one fuel molecule abstract H atoms from the preliminary radicals generated from the other fuel species [71,72]. It has been mentioned in a recent work by Tripathi et al. [73] that the co-oxidation reactions may be considered insignificant if the fuel molecules are decomposing at different temperature zones. In order to investigate this fact, the percentage of unburnt isooctane and 2MF with respect to their initial amount is plotted across the flame coordinate in Fig. 7 for stoichiometric 80% isooctane/20% 2MF blend in air at 333 K unburnt gas temperature and atmospheric pressure. Furthermore, the corresponding flame temperature in the same computational domain is shown in the secondary axis. It may be seen from the figure that the 2MF molecule breaks more readily than isooctane in the initial part of the preheat zone. This is due to the lower minimum

C—H bond dissociation energy in 2MF structure (306.9 kJ/mol) [25] than isooctane (389.69 kJ/mol) [74]. The maximum difference in unburnt 2MF and isooctane fractions is observed at a temperature of 880 K. This difference may be attributed to the NTC behaviour of the isooctane combustion chemistry. As discussed earlier, the $HO_2$ radicals dominate the combustion kinetics in the NTC region. These radicals form the hydrogen peroxide ($H_2O_2$) molecule through the reaction $HO_2+C_8H_{18} = H_2O_2+C_8H_{17}$. The $H_2O_2$ is a metastable species that does not break according to the chain branching reaction $H_2O_2(+M) = OH + OH(+M)$ unless the temperature goes high enough. Therefore, the consumption rate of the isooctane molecule is decelerated in the NTC region due to the formation of the $H_2O_2$ molecules [59]. In order to identify the dominance of the NTC chemistry, the variation of the mole fraction of $H_2O_2$ ($Y_{H2O2}$) in ppm across the flame coordinate is embedded in Fig. 7. It may be seen from the figure that the peak value of $H_2O_2$ mole fraction also occurs at 880 K. Beyond this point as the temperature increases, the above-mentioned chain branching breakup of the $H_2O_2$ molecule gains relevance. Therefore, the slope of the unburnt isooctane percentage in Fig. 7 becomes steeper. This faster depletion of isooctane results in the unburnt fractions of both the fuel components to be nearly equal ($\approx$ 16%) at a temperature of 1284 K. Again, beyond this point of inflection, the consumption rate of both isooctane and 2MF slows down. It is also clear from the temperature curve that this inflection point is the juncture between the preheat and reaction zones where the ignition occurs.

The main reason behind this reduction in reactivity of the fuel components may be derived from the rate of production analysis shown in Fig. 8 (a) and (b) for isooctane and 2MF respectively. It may be seen from Fig. 8 (a) that the isooctane ($IC_8H_{18}$) consumption via H abstraction reactions to produce four isomers of isooctyl radicals ($AC_8H_{17}$, $BC_8H_{17}$, $CC_8H_{17}$ and $DC_8H_{17}$) [58,75] attains its peak at the point of inflection. A similar observation has been reported earlier by Bhattacharya et al. [76] as well. On the other hand, the consumption of the isooctane molecule is only governed by the unimolecular decomposition route $IC_8H_{18}=YC_7H_{15}+CH_3$ at temperatures higher than 1284 K beyond the ignition point. Hence, it may be inferred that the H abstraction reactions accelerate the isooctane fuel consumption in the preheat zone and ignition occurs when such consumption reaches its peak. On a similar note, the consumption of the 2MF fuel is also partially dominated by the H abstraction reactions 2MF+OH=MF22J+H$_2$O and 2MF+H=MF22J+H$_2$ in the preheat zone of the flame up to the inflection point (Fig. 8 (b)). In addition to these, the ring opening reactions 2MF+H=CH$_2$CHCHCO+CH$_3$ and 2MF+H=P4E2O3J-C1 also contribute significantly towards the consumption of the 2MF molecule in the preheat zone. It may be seen that the consumption of 2MF takes place in two stages through the reaction 2MF+H=P4E2O3J-C1. However, the consumption rate decelerates abruptly beyond the ignition point as all of these main decomposition routes are slowed down.

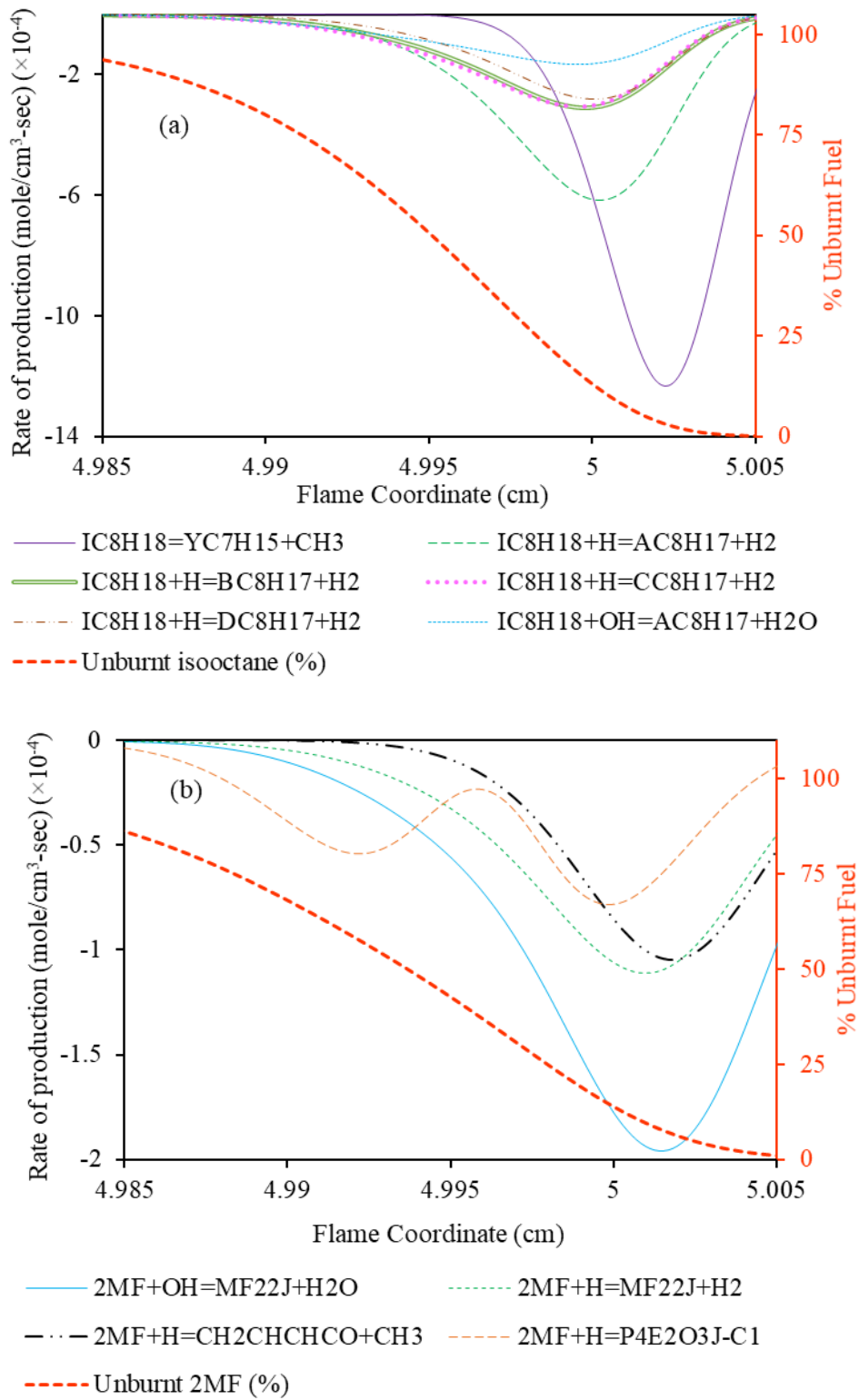

Figure 8. Variation of the unburnt fuel fraction and rate of consumption of the fuel molecule by individual reactions across the computational domain for (a) isooctane and (b) 2MF.

As seen from the above discussion, isooctane and 2MF decompose at different rates at different regions of the flame involving isooctane/2MF blend in air. Furthermore, the peak rates of the decomposition reactions occur at different positions in the preheat zone. Therefore, the possibility of co-oxidation reactions between the radicals directly generated from the fuel molecule breakup should be highest in this zone. In order to assess the chances of such reactions, the mole fraction (in ppm) of the primary decomposition species of isooctane (i.e. the isooctyl radicals) and 2MF (i.e. 2-furylmethyl (MF22J), 1-oxo-1,3-butadiene or vinyl ketene ($CH_2CHCHCO$) and 3,4-pentadiene-1-one-2-yl (P4E2O3J-C1)) are shown in Fig. 9. It may be seen from the figure that all the isooctyl radicals show their peak in the initial part of the preheat zone while the primary decomposition radicals from 2MF are distributed over a wider length of the flame coordinate downstream. The relative positions of these heavy radicals from 2MF and isooctane contradict the observation made in Fig. 7 where the 2MF molecule is seen to be decomposing earlier than isooctane. Therefore, it may be inferred that the preferential mass diffusion of the primary decomposition radicals determines their relative position in the flame. Furthermore, the peak mole fractions of MF22J, $CH_2CHCHCO$ and P4E2O3J-C1 are a few orders of magnitude higher than the four isomers of the isooctyl radicals despite the fact that that fuel stream contains isooctane and 2MF in 4:1 ratio.

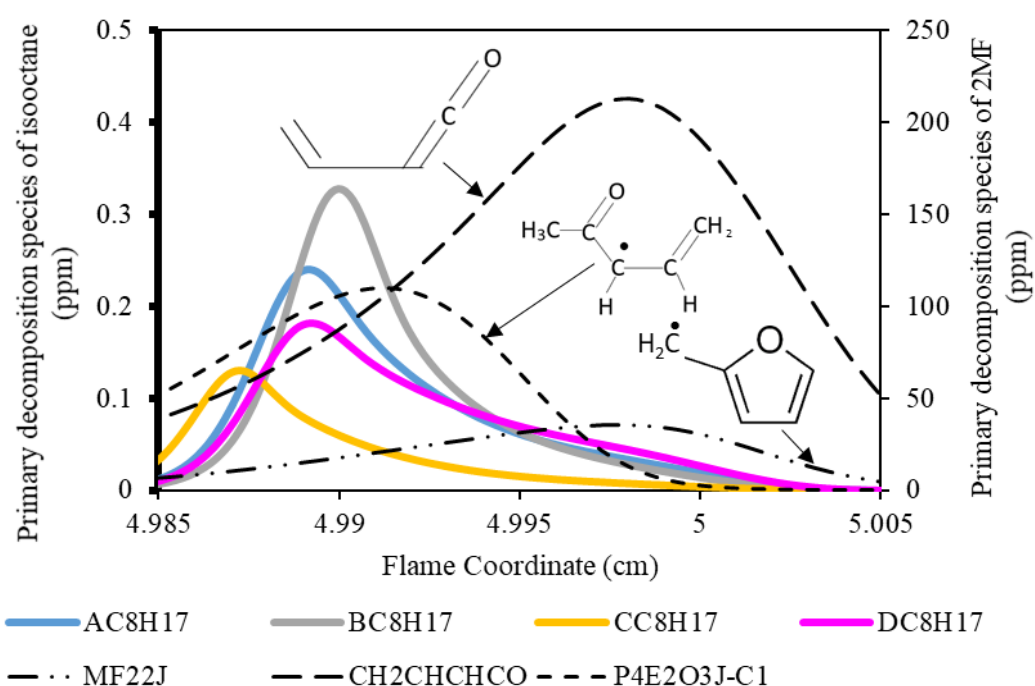

Figure 9. Distribution of primary decomposition radicals from isooctane and 2MF across the computational domain for stoichiometric 80% isooctane/20% 2MF blend in air at 333 K unburnt gas temperature and 1 atm pressure.

In order to investigate this disparity further, the rate of production of these radicals have been analysed at the point of their respective peaks. As all the isooctyl radicals follow similar chemical path, the rate of production of one of them ($BC_8H_{17}$) with the highest presence in the flame is shown in Fig. 10 (a). as seen from the figure, these heavy radicals are mainly produced from the fuel ($IC_8H_{18}$) molecule through H abstraction by H and OH radicals. Then, these isooctyl radicals are primarily consumed through the β-scission breakup like $BC_8H_{17}=YC_7H_{14}+CH_3$. On the other hand, $CH_2CHCHCO$ is generated through both the paths $2MF+H=CH_2CHCHCO+CH_3$ and $P4E2O3J\text{-}C1=CH_3+CH_2CHCHCO$ (Fig. 10 (b)). It may be mentioned in this context that the P4E2O3J-C1 radical is generated through the routes $2MF+H=P4E2O3J\text{-}C1$ and $P4E2O3J\text{-}C2=P4E2O3J\text{-}C1$ where P4E2O3J-C2 is the secondary conformer of P4E2O3J-C1. These multiple generation paths explain the higher amount of $CH_2CHCHCO$ and P4E2O3J-C1 in the flame compared to the isooctyl radicals in Fig. 9. The 2-furylmethyl radical is generated through the H abstraction from the 2MF structure by H and OH radicals (Fig. 10 (c)). Furthermore, a large portion of the MF22J combines with the $CH_3$ radicals to form 2-ethylfuran (E2F). Therefore, the mole fraction of MF22J is significantly low in comparison to $CH_2CHCHCO$ and P4E2O3J-C1 in Fig. 9.

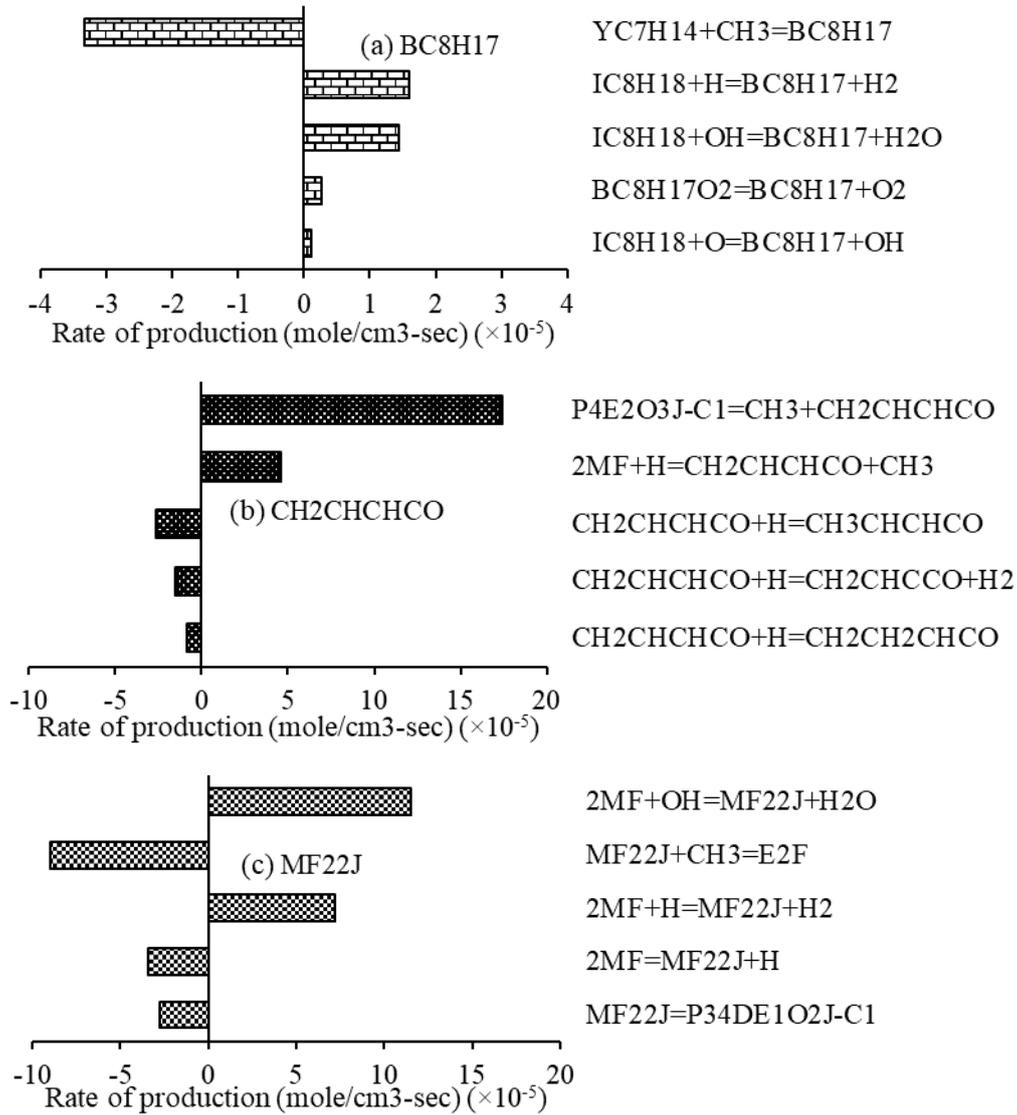

Figure 10. Rate of production analysis of (a) BC$_8$H$_{17}$, (b) CH$_2$CHCHCO and (c) MF22J.

It may be summarized from the above discussion that the decomposition of the fuel molecules occurs in a spatially distributed manner inside the premixed flame involving 80% isooctane/20% 2MF blend in air at stoichiometric condition. Furthermore, there is significant inhomogeneity in the distribution of initial heavy species generated from both the fuel molecule breakups due to preferential diffusion. The peak mole fractions of the primary decomposition species also vary by several orders of magnitude for isooctane and 2MF. The combination of all these factors minimize the possibilities of co-oxidation reactions among these heavy species. Therefore, such reactions have not been considered in the present model.

## 4.2 Effects of blending 2MF with isooctane on laminar burning velocity, soot precursors and NO formation

The laminar flame discussed in the above section is of great practical significance. An increase in the mass burning flux associated with such a laminar flame increases the flame stability in a real-world combustor for a particular fuel blend. Moreover, augmented mass burning flux—or laminar burning velocity—brings the thermal efficiency of the SI engine close to its air standard limit [47,77,78]. It may be seen from Fig. 11 that the peak value of $S_L$ increases by around 8.6% when the 2MF mole fraction is increased from 10% to 90% in the isooctane/2MF binary fuel blend at $\varphi = 1.1$ for unburnt gas pressure and temperature of 1 atm and 333 K respectively. Furthermore, the difference in $S_L$ between the two binary blends becomes increasingly conspicuous as the equivalence ratio increases. In order to investigate the increase in $S_L$ due to 2MF addition in isooctane, a sensitivity analysis was carried out at stoichiometric condition. The results are shown in Fig. 12. The sensitivity analysis suggests that the key influencing reactions are same for both the fuel blends. However, there are quantitative differences among the intensities of such influences of individual reactions. The positive contribution of the CO oxidation reaction $CO+OH=CO_2+H$ increases with the increase in 2MF amount in fuel. On the other hand, the reactions like $H_2O+M=H+OH+M$, $H+CH_3(+M)= CH_4(+M)$, $HCO+OH=CO+H_2O$ and $IC_4H_8=IC_4H_7+H$ whose forward progress reduces the value of $S_L$ show lesser sensitivity with the increase in 2MF amount in the binary fuel blend. These chemical factors combine to increase the laminar burning velocity due to 2MF addition in isooctane.

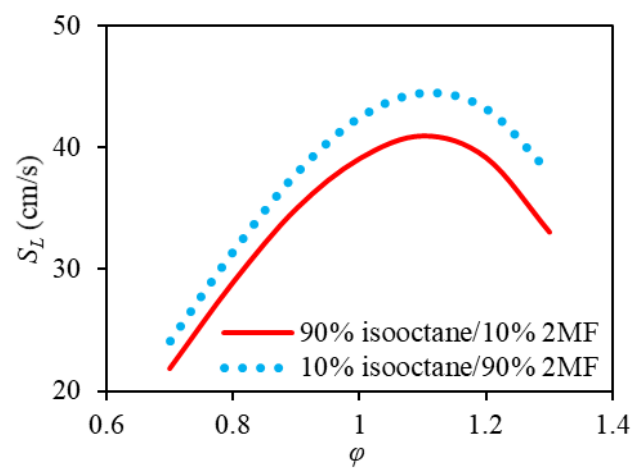

Figure 11. Variation of laminar burning velocity with equivalence ratio for 90% isooctane/10% 2MF and 10% isooctane/90% 2MF blends in air at atmospheric pressure and unburnt gas temperature of 333 K.

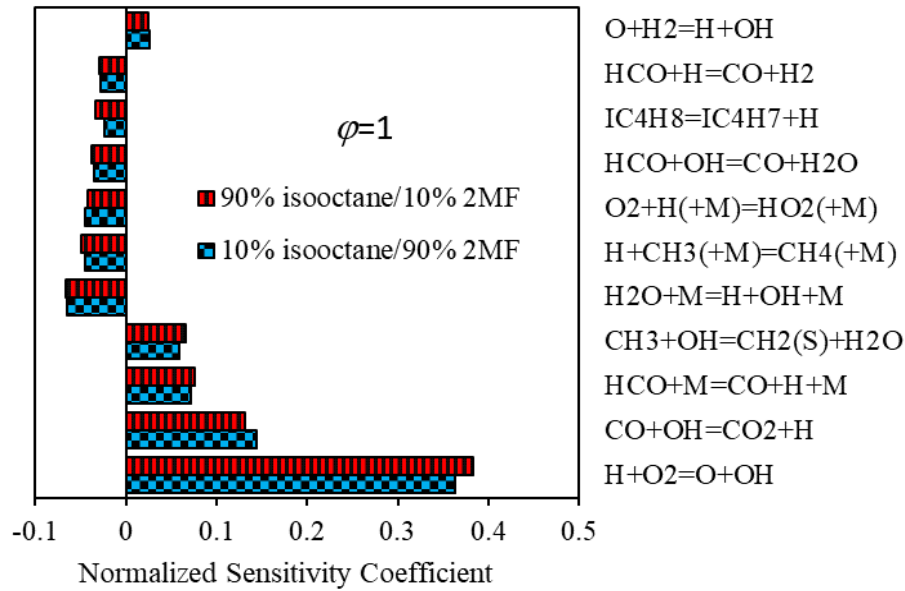

Figure 12. Sensitivity analysis on the laminar burning velocity of 90% isooctane/10% 2MF and 10% isooctane/90% 2MF blends in air at atmospheric pressure, unburnt gas temperature of 333 K and $\varphi = 1$.

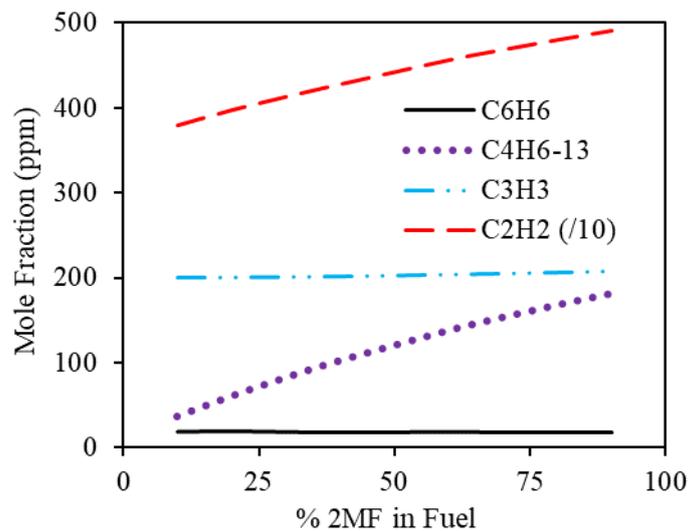

Figure 13. Variation of peak soot precursor mole fraction (ppm) in the flame for different amounts of 2MF in isooctane/2MF blend in air at atmospheric pressure, unburnt gas temperature of 333 K and $\varphi = 1.3$.

The above discussion reveals that the addition of 2MF to isooctane is advantageous from the perspective of engine efficiency and flame stability inside the combustor. However, along with the improvement in performance, the effects of blending 2MF to isooctane on the emission of different pollutants (like soot and NO) needs to be assessed as well. While soot is majorly emitted from rich flames, the NO is one of the main pollutants from the near stoichiometric flames that have high flame temperatures. It is widely believed that the polycyclic aromatic hydrocarbons (PAHs) are the gaseous precursors of soot particles generated from a flame. Furthermore, the hydrogen abstraction acetylene addition (HACA) to the benzene ($C_6H_6$) ring [79,80] is a widely accepted route for the growth of PAHs. The formation of the initial $C_6H_6$ molecule involves reactions between certain smaller molecules (like acetylene ($C_2H_2$), propargyl ($C_3H_3$) and 1,3-butadiene ($C_4H_6$-13) [81,82]) as well. Therefore, the peak mole fractions (in ppm) of $C_2H_2$, $C_3H_3$, $C_4H_6$-13 and $C_6H_6$ in the premixed flames of isooctane/2MF blends in air are plotted in Fig. 13 as a function of 2MF mole fraction (in %) in the binary fuel blend for $\varphi = 1.3$. It may be seen from the figure that the amount of $C_3H_3$ and $C_6H_6$ are almost unaltered with the increase in the 2MF mole fraction. This is due to the fact that the chain terminating reaction $C_3H_3+C_3H_3=C_6H_6$ majorly controls the conversion of $C_3H_3$ to $C_6H_6$ [82]. On the other hand, the increase for $C_2H_2$ and $C_4H_6$-13 is around 30% and 390% respectively with the increase in the 2MF mole fraction from 10% to 90% in the binary isooctane/2MF blend.

In order to investigate these facts further, the reaction paths leading to $C_2H_2$ and $C_4H_6$-13 from isooctane and 2MF molecule has been analysed in Fig. 14 (a) and (b) respectively for the 80% isooctane/20% 2MF blend flame. The reaction paths for $C_2H_2$ and $C_4H_6$-13 have been drawn at the locations of their respective peak mole fraction. It may be seen from Fig. 14 (a) that the $C_2H_2$ molecule is generated through the route $IC_8H_{18} \rightarrow$ methyl ($CH_3$) $\rightarrow$ ethane($C_2H_6$) $\rightarrow$ ethyl($C_2H_5$) $\rightarrow$ ethene($C_2H_4$) $\rightarrow$ vinyl($C_2H_3$) $\rightarrow$ acetylene($C_2H_2$). The $CH_3$ radical is generated from the thermal decomposition of isooctane through the reaction $IC_8H_{18}= yC_7H_{15}+CH_3$. Furthermore, the β-scission of the 2,4-dimethyl-propan-2-yl radical ($yC_7H_{15}$) initiates the path towards the formation of $C_4H_6$-13. Some of the key intermediates during the formation of $C_4H_6$-13 include allene ($aC_3H_4$), propyne ($pC_3H_4$), propargyl ($C_3H_3$) and 1,2-butadiene ($C_4H_6$-12). On the other hand, it is evident from Fig. 14 (b) that $C_4H_6$-13 is generated from 2MF molecule through a much shorter path with only two isomeric intermediates (viz. 1-buten-1-yl ($C_4H_7$-11) and 1-Methylallyl ($C_4H_7$-13) radicals). Therefore, this shorter path contributes heavily towards the formation of $C_4H_6$-13 in the premixed flame when 2MF is added to isooctane. The $C_2H_2$ molecule is generated from the initial breakup mechanism of the 2MF molecule through the path 2MF → 2-furanylmethyl (MF22J) → 3,4-pentadiene-1-one-2-yl (P34DE1O2J-C1) → secondary conformer of 3,4-pentadiene-1-one-2-yl (P34DE1O2J-C2) → 1,4-pentadiene-1-one-3-yl (P14DE1O3J-C1) → 2,4-pentadiene-1-one-1-yl (P24DE1O1J) → 1,3-butdien-4-yl ($C_4H_5$-N) → acetylene ($C_2H_2$). It may be noted that there is little contribution of $IC_8H_{18}$ on the $C_2H_2$ generation path from 2MF and vice versa. Therefore, as a combination of these independent formation paths originating from the fuel components, the mole fraction of $C_2H_2$ increases proportionately with the 2MF mole fraction in $IC_8H_{18}$/2MF binary fuel blend as seen in Fig. 13.

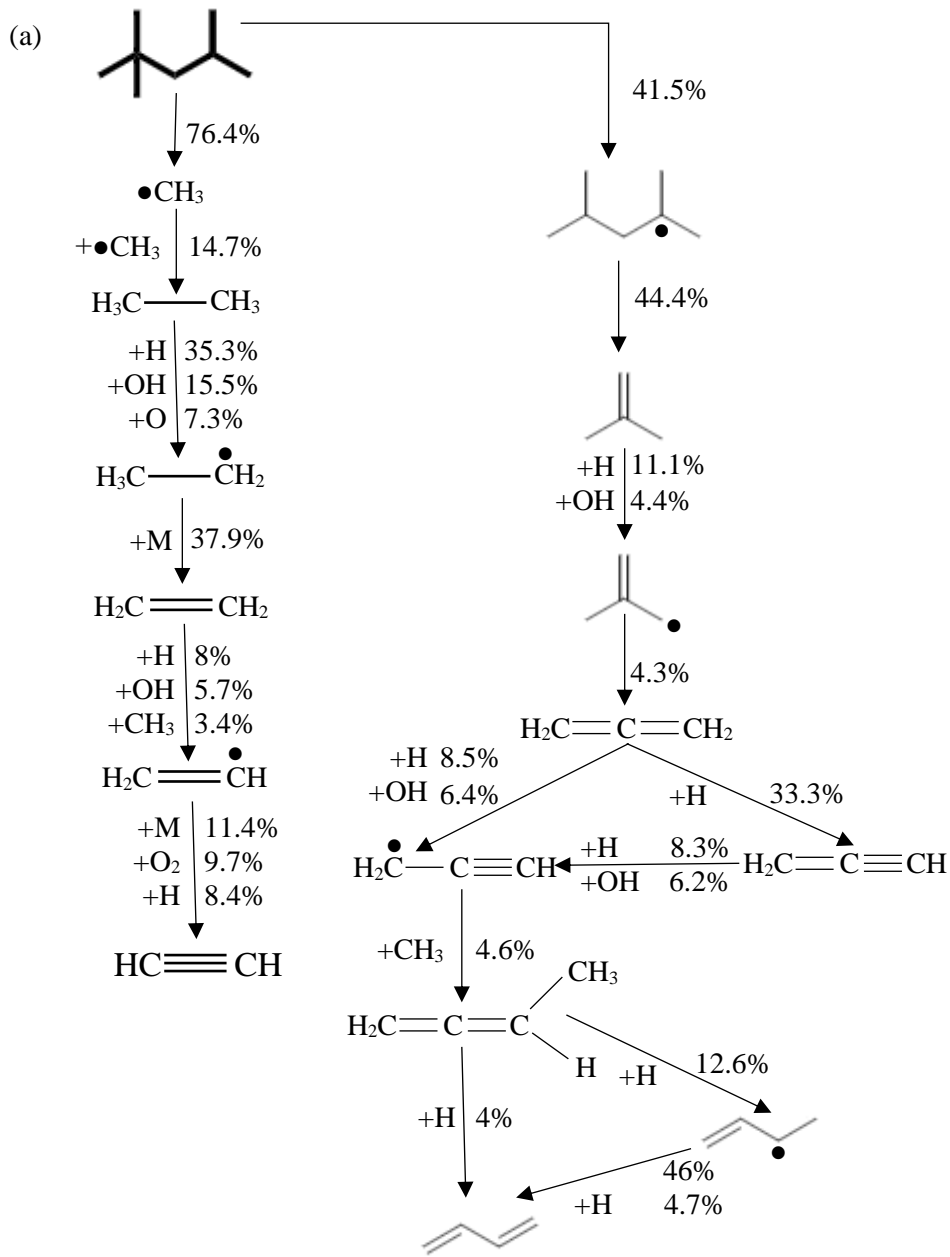

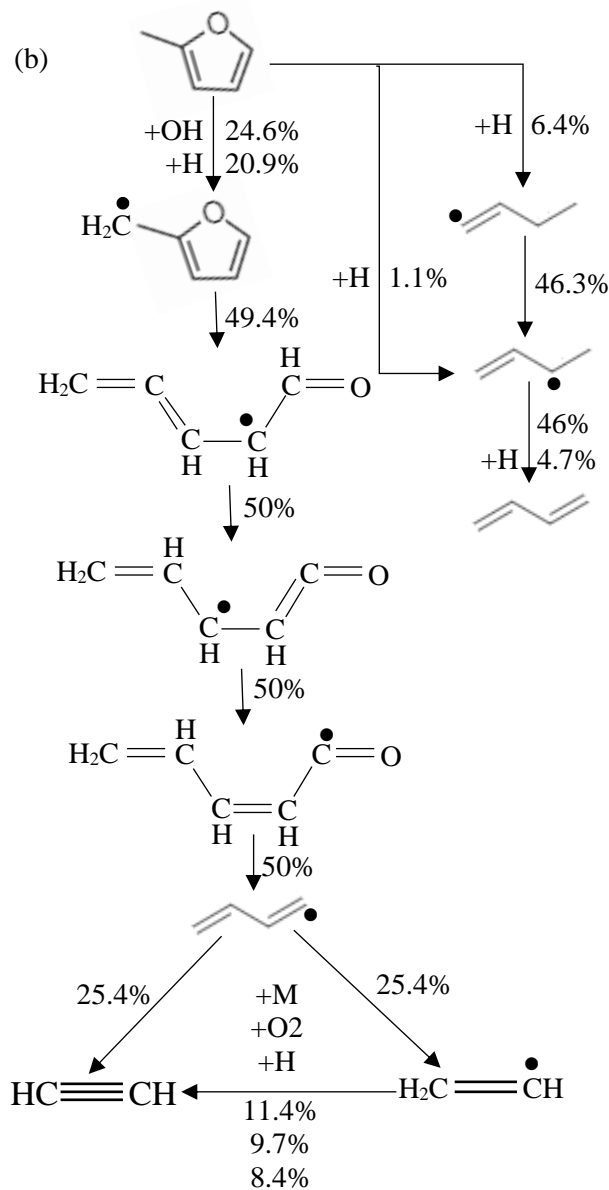

Figure 14. Reaction path showing the formation of acetylene ($C_2H_2$) and 1,3-butadiene ($C_4H_6$-13) from (a) isooctane ($IC_8H_{18}$) and (b) 2MF molecule. The flame considered involves 80% isooctane/20% 2MF blend in air at atmospheric pressure, unburnt gas temperature of 333 K and $\varphi = 1.3$.

After analysing the formation of soot precursors, the effect of 2MF addition to isooctane on the formation of NO has been assessed. It is worth mentioning in this regard that the NO can be produced through either the prompt or the thermal route in case the fuel itself does not contain nitrogen. While the thermal route involving the Zeldovich mechanism [83] is governed by the flame temperature, the prompt route involves the interactions between the hydrocarbon and nitrogen molecules and is mostly active in the low temperature rich flames. A recent review work by Glarborg et al. [84] suggests that there is still huge scope for improvement in the understanding of nitrogen chemistry from combustion perspective. Hence, the present work aims to propose a chemical kinetic mechanism capable of predicting the $NO_x$ emissions from flames involving isooctane/2MF blends. However, it may be seen from the inset in Fig. 15 that the $NO_x$ sub-mechanism of Alexandrino et al. [49] in the present model provides acceptable qualitative prediction of NO emission in the range $0.75 \leq \varphi \leq 1.07$ only. The experimental dataset used for this validation has been taken from the work of Bartok et al.[85] involving atmospheric pressure methane/air combustion in a well stirred reactor for initial temperature of 463.7 K and 3 ms residence time. Therefore, the effect of 2MF addition to isooctane towards NO formation in premixed flame is shown in Fig. 15 at atmospheric pressure and stoichiometric condition. The unburnt gas temperature is 333 K. It may be seen from the figure that the peak NO mole fraction (in ppm) increases by around 21% as the 2MF mole fraction increases from 10% to 90% in the isooctane/2MF blend. This increase may be attributed to the increase in flame temperature and domination of the Zeldovich mechanism due to the addition of 2MF as shown in the secondary axis of the figure.

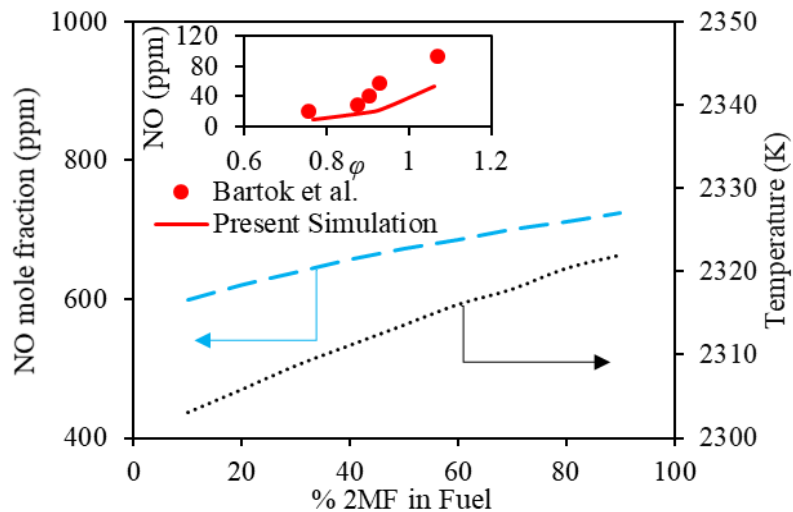

Figure 15. Variation of peak NO mole fraction (ppm) and temperature in the flame for different amounts of 2MF in isooctane/2MF blend in air at atmospheric pressure, unburnt gas temperature of 333 K and $\varphi = 1$. The inset shows the validation of the NO mechanism in the present model against the experimental data from Bartok et al. involving atmospheric pressure methane/air combustion in a well stirred reactor for initial temperature of 463.7 K and 3 ms residence time.

## 5. Conclusions

In the present work, a skeletal chemical kinetic mechanism consisting of 252 species and 1288 reactions has been proposed for the simulation of premixed flames involving isooctane/2MF blends. The mechanism has been validated against a wide range of experimental data on ignition delay time, laminar burning velocity and species profiles from burner stabilized premixed flames and well stirred reactors for the individual fuel components as well as their blends. The following conclusions can be drawn from the present study:

1) The likelihood of co-oxidation reactions among the species generated from the initial decomposition of the fuel molecules is minimized due to the following reasons:
   a) Contrary to the homogeneous distribution of the reactants during the measurement of ignition delay time and well stirred reactor experiments, the decomposition of the fuel molecules occurs in a spatially distributed manner inside the premixed flames.

b) Similarly, the preferential diffusion of these primary decomposition species leads to an inhomogeneous distribution to prevail in the flame.

c) The mole fractions of the primary decomposition species have also been found to differ by several orders of magnitude for isooctane and 2MF.

2) The difference in the laminar burning velocity between 90% isooctane/10% 2MF and 10% isooctane/90% binary fuel blend increases with the increase in equivalence ratio. Furthermore, the difference in peak value of the laminar burning velocity is around 8.6% between these two blends at $\varphi = 1.1$. This marginal variation is advantageous from the SI engine application perspective as this would entail nominal design modifications.

3) As far as the gaseous soot precursors are concerned, it has been observed in the present work that the addition of 2MF to isooctane has little influence on the amount of peak benzene ($C_6H_6$) and propargyl ($C_3H_3$) mole fractions in the premixed flame corresponding to $\varphi = 1.3$. However, the peak mole fraction of acetylene ($C_2H_2$) and 1,3-butadiene ($C_4H_6$-13) in the flame increases with the increase in 2MF mole fraction in the isooctane/2MF blend. The primary cause behind such rise is the generation paths leading to $C_2H_2$ and $C_4H_6$-13 from the 2MF molecule which is unaffected by the presence of isooctane.

4) On the other hand, the $NO_x$ sub-mechanism in the present model provides acceptable prediction of NO emission in the range $0.75 \leq \varphi \leq 1.07$ only. In this region, the NO production is primarily governed by the thermal route, i.e. the Zeldovich mechanism. It has been observed in the present study that the addition of 2MF in isooctane increases the flame temperature, thereby increasing the NO mole fraction.

5) Finally, the present study suggests that the prompt NO mechanism of isooctane/2MF blend needs to be further improved. Furthermore, there is scope for the evaluation of the contribution of co-oxidation reactions between the heavy molecules generated from the fuels from a quantitative perspective in future.

**Acknowledgement**

This research did not receive any specific grant from funding agencies in the public, commercial, or not-for-profit sectors.

**References**

[1]   BP Energy. 2018 BP Energy Outlook. 2018. doi:10.1088/1757-899X/342/1/012091.

[2]   Lange JP, Van Der Heide E, Van Buijtenen J, Price R. Furfural-A promising platform for lignocellulosic biofuels. ChemSusChem 2012;5:150–66. doi:10.1002/cssc.201100648.

[3]   Zhong S, Daniel R, Xu H, Zhang J, Turner D, Wyszynski ML, et al. Combustion and emissions of 2,5-dimethylfuran in a direct-injection spark-ignition engine. Energy and Fuels 2010;24:2891–2899. doi:10.1021/ef901575a.

[4]   Jężak S, Dzida M, Zorębski M. High pressure physicochemical properties of 2-methylfuran and 2,5-dimethylfuran – second generation biofuels. Fuel 2016;184:334–343. doi:10.1016/j.fuel.2016.07.025.

[5]   Leitner W, Klankermayer J, Pischinger S, Pitsch H, Kohse-Höinghaus K. Advanced Biofuels and Beyond: Chemistry Solutions for Propulsion and Production. Angew Chemie - Int Ed 2017;56:5412–52. doi:10.1002/anie.201607257.

[6]   Tran LS, Sirjean B, Glaude PA, Kohse-Höinghaus K, Battin-Leclerc F. Influence of substituted furans on the formation of Polycyclic Aromatic Hydrocarbons in flames. Proc Combust Inst 2015;35:1735–1743. doi:10.1016/j.proci.2014.06.137.

[7]   Thewes M, Muether M, Pischinger S, Budde M, Brunn A, Sehr A, et al. Analysis of the impact of 2-methylfuran on mixture formation and combustion in a direct-injection spark-ignition engine. Energy and Fuels 2011;25:5549–5561. doi:10.1021/ef201021a.

[8]   Hoppe F, Heuser B, Thewes M, Kremer F, Pischinger S, Dahmen M, et al. Tailor-made fuels for future engine concepts. Int J Engine Res 2016;17:16–27. doi:10.1177/1468087415603005.

[9]   Hoppe F, Burke U, Thewes M, Heufer A, Kremer F, Pischinger S. Tailor-Made Fuels from Biomass: Potentials of 2-butanone and 2-methylfuran in direct injection spark ignition engines. Fuel 2016;167:106–17. doi:10.1016/j.fuel.2015.11.039.

[10]  Wang C, Xu H, Daniel R, Ghafourian A, Herreros JM, Shuai S, et al. Combustion characteristics and emissions of 2-methylfuran compared to 2,5-dimethylfuran, gasoline and ethanol in a DISI engine. Fuel 2013;103:200–11. doi:10.1016/j.fuel.2012.05.043.

[11]  Chinnathambi P, Wadkar C, Toulson E. Impact of CO2 Dilution on Ignition Delay Times of Iso-Octane at 15% and 30% Dilution Levels in a Rapid Compression Machine. SAE Tech. Pap. Ser., 2019, p. 2019-01–0569. doi:10.4271/2019-01-0569.

[12]  Mansfield AB, Wooldridge MS, Di H, He X. Low-temperature ignition behavior of iso-octane. Fuel 2015;139:79–86. doi:10.1016/j.fuel.2014.08.019.

[13]  Baloo M, Dariani BM, Akhlaghi M, Chitsaz I. Effect of iso-octane/methane blend on laminar burning velocity and flame instability. Fuel 2015;144:264–73. doi:10.1016/j.fuel.2014.11.043.

[14]  Baloo M, Dariani BM, Akhlaghi M, Aghamirsalim M. Effects of pressure and temperature on laminar burning velocity and flame instability of iso-octane/methane fuel blend. Fuel 2016;170:235–44. doi:10.1016/j.fuel.2015.12.039.

[15]  Jouzdani S, Eldeeb MA, Zhang L, Akih-Kumgeh B. High-Temperature Study of 2-Methyl Furan and 2-Methyl Tetrahydrofuran Combustion. Int J Chem Kinet 2016;48:491–503. doi:10.1002/kin.21008.

[16]  Wei L, Tang C, Man X, Huang Z. Shock-tube experiments and kinetic modeling of 2-methylfuran ignition at elevated pressure. Energy and Fuels 2013;27:7809–16. doi:10.1021/ef401809y.

[17] Kim G, Almansour B, Park S, Terracciano A, Vasu S, Zhang K, et al. Laminar Burning Velocities of High-Performance Fuels Relevant to the Co-Optima Initiative. SAE Tech. Pap., 2019, p. 2019-01–0571. doi:10.4271/2019-01-0571.

[18] Sarathy SM, Farooq A, Kalghatgi GT. Recent progress in gasoline surrogate fuels. Prog Energy Combust Sci 2018;65:67–108. doi:10.1016/j.pecs.2017.09.004.

[19] Zhen X, Wang Y, Liu D. An overview of the chemical reaction mechanisms for gasoline surrogate fuels. Appl Therm Eng 2017;124:1257–1268. doi:10.1016/j.applthermaleng.2017.06.101.

[20] Bhattacharya A, Banerjee DK, Mamaikin D, Datta A, Wensing M. Effects of Exhaust Gas Dilution on the Laminar Burning Velocity of Real-World Gasoline Fuel Flame in Air. Energy & Fuels 2015;29:6768–79. doi:10.1021/acs.energyfuels.5b01299.

[21] Huang Y, Sung CJ, Eng JA. Laminar flame speeds of primary reference fuels and reformer gas mixtures. Combust Flame 2004;139:239–51. doi:10.1016/j.combustflame.2004.08.011.

[22] Li R, He G, Zhang D, Qin F. Skeletal Kinetic Mechanism Generation and Uncertainty Analysis for Combustion of Iso-octane at High Temperatures. Energy and Fuels 2018;32:3842–3850. doi:10.1021/acs.energyfuels.7b02838.

[23] Basevich VY, Belyaev AA, Medvedev SN, Frolov SM, Frolov FS. A Detailed Kinetic Mechanism of Multistage Oxidation and Combustion of Octanes. Russ J Phys Chem B 2018;12:448–57. doi:10.1134/s1990793118030223.

[24] Wang Q, Tang X, Yang J, Zhai Y, Zhang Y, Cao C, et al. Investigations on Pyrolysis of Isooctane at Low and Atmospheric Pressures. Energy and Fuels 2019;33:3518−3528. doi:10.1021/acs.energyfuels.8b04029.

[25] Simmie JM, Curran HJ. Formation enthalpies and bond dissociation energies of alkylfurans. the strongest C~X bonds known? J Phys Chem A 2009;113:5128–37. doi:10.1021/jp810315n.

[26] Somers KP, Simmie JM, Gillespie F, Burke U, Connolly J, Metcalfe WK, et al. A high temperature and atmospheric pressure experimental and detailed chemical kinetic modelling study of 2-methyl furan oxidation. Proc Combust Inst 2013;34:225–32. doi:10.1016/j.proci.2012.06.113.

[27] de Goey LPH, van Maaren A, Ouax RM. Stabilization of Adiabatic Premixed Laminar Flames on a Flat Flame Burner. Combust Sci Technol 1993;92:201–7. doi:10.1080/00102209308907668.

[28] Somers KP, Simmie JM, Metcalfe WK, Curran HJ. The pyrolysis of 2-methylfuran: A quantum chemical, statistical rate theory and kinetic modelling study. Phys Chem Chem Phys 2014;16:5349–67. doi:10.1039/c3cp54915a.

[29] Cheng Z, Niu Q, Wang Z, Jin H, Chen G, Yao M, et al. Experimental and kinetic modeling studies of low-pressure premixed laminar 2-methylfuran flames. Proc Combust Inst 2017;36:1295–302. doi:10.1016/j.proci.2016.07.032.

[30] Liu D, Togbé C, Tran LS, Felsmann D, Oßwald P, Nau P, et al. Combustion chemistry and flame structure of furan group biofuels using molecular-beam mass spectrometry and gas chromatography - Part I: Furan. Combust Flame 2014;161:748–65. doi:10.1016/j.combustflame.2013.05.028.

[31] Tran L-S, Togbé C, Liu D, Felsmann D, Oßwald P, Glaude P-A, et al. Combustion chemistry and flame structure of furan group biofuels using molecular-beam mass spectrometry and gas chromatography - Part II: 2-Methylfuran. Combust Flame 2014;161:766–779. doi:10.1016/j.combustflame.2013.05.027.

[32] Togbé C, Tran LS, Liu D, Felsmann D, Oßwald P, Glaude PA, et al. Combustion chemistry and flame structure of furan group biofuels using molecular-beam mass spectrometry and gas chromatography - Part III: 2,5-Dimethylfuran. Combust Flame 2014;161:780–797. doi:10.1016/j.combustflame.2013.05.026.

[33] Sirjean B, Fournet R, Glaude PA, Battin-Leclerc F, Wang W, Oehlschlaeger MA. Shock tube and chemical kinetic modeling study of the oxidation of 2,5-dimethylfuran. J Phys Chem A 2013;117:1371−1392. doi:10.1021/jp308901q.

[34] Tran LS, Wang Z, Carstensen HH, Hemken C, Battin-Leclerc F, Kohse-Höinghaus K. Comparative experimental and modeling study of the low- to moderate-temperature oxidation chemistry of 2,5-dimethylfuran, 2-methylfuran, and furan. Combust Flame 2017;181:251–69. doi:10.1016/j.combustflame.2017.03.030.

[35] Weber I, Friese P, Olzmann M. H-Atom-Forming Reaction Pathways in the Pyrolysis of Furan, 2-Methylfuran, and 2,5-Dimethylfuran: A Shock-Tube and Modeling Study. J Phys Chem A 2018;122:6500–8. doi:10.1021/acs.jpca.8b05346.

[36] Lu T, Law CK. Toward accommodating realistic fuel chemistry in large-scale computations. Prog Energy Combust Sci 2009;35:192–215. doi:10.1016/j.pecs.2008.10.002.

[37] Pope SB. Small scales, many species and the manifold challenges of turbulent combustion. Proc Combust Inst 2013;32:1527–1535. doi:10.1016/j.proci.2012.09.009.

[38] Pope SB, Ren Z. Efficient implementation of chemistry in computational combustion. Flow, Turbul Combust 2009;82:437–453. doi:10.1007/s10494-008-9145-3.

[39] Tomlin AS, Turányi T, Pilling MJ. Mathematical tools for the construction, investigation and reduction of combustion mechanisms. Compr. Chem. Kinet., 1997, p. 293–437. doi:10.1016/S0069-8040(97)80019-2.

[40] Peters N, Rogg B. Reduced Kinetic Mechanisms for Applications in Combustion Systems. 1993. doi:10.1007/978-3-662-29391-1_1.

[41] Ren Z, Pope SB, Vladimirsky A, Guckenheimer JM. The invariant constrained equilibrium edge preimage curve method for the dimension reduction of chemical kinetics. J Chem Phys 2006;124:114111. doi:10.1063/1.2177243.

[42] Pope SB. Computationally efficient implementation of combustion chemistry using in situ adaptive tabulation. Combust Theory Model 1997;1:41–63. doi:10.1080/713665229.

[43] Oijen JA van, Goey LPH de. Modelling of Premixed Laminar Flames using Flamelet-Generated Manifolds. Combust Sci Technol 2000;161:113–37. doi:10.1080/00102200008935814.

[44] Xie W, Lu Z, Ren Z. Rate-controlled constrained equilibrium for large hydrocarbon fuels with NTC. Combust Theory Model 2019;23:226–44. doi:10.1080/13647830.2018.1513566.

[45] Somers KP, Simmie JM, Gillespie F, Conroy C, Black G, Metcalfe WK, et al. A comprehensive experimental and detailed chemical kinetic modelling study of 2,5-dimethylfuran pyrolysis and oxidation. Combust Flame 2013;160:2291–2318. doi:10.1016/j.combustflame.2013.06.007.

[46] Pepiot-Desjardins P, Pitsch H. An efficient error-propagation-based reduction method for large chemical kinetic mechanisms. Combust Flame 2008;154:67–81. doi:10.1016/j.combustflame.2007.10.020.

[47] Bhattacharya A, Datta A. Effects of blending 2,5-dimethylfuran on the laminar burning velocity and ignition delay time of isooctane/air mixture. Combust Theory Model 2018:1–22. doi:10.1080/13647830.2018.1492153.

[48] Yoo CS, Luo Z, Lu T, Kim H, Chen JH. A DNS study of ignition characteristics of a lean iso-octane/air mixture under HCCI and SACI conditions. Proc Combust Inst 2013;34:2985–93. doi:10.1016/j.proci.2012.05.019.

[49] Alexandrino K, Millera Á, Bilbao R, Alzueta MU. 2-methylfuran Oxidation in the Absence and Presence of NO. Flow, Turbul Combust 2016;96:343–62. doi:10.1007/s10494-015-9635-z.

[50] Ranzi E, Frassoldati A, Stagni A, Pelucchi M, Cuoci A, Faravelli T. Reduced kinetic schemes of complex reaction systems: Fossil and biomass-derived transportation fuels. vol. 46. 2014. doi:10.1002/kin.20867.

[51] Ranzi E, Frassoldati A, Grana R, Cuoci A, Faravelli T, Kelley AP, et al. Hierarchical and comparative kinetic modeling of laminar flame speeds of hydrocarbon and oxygenated fuels. Prog Energy Combust Sci 2012;38:468–501. doi:10.1016/j.pecs.2012.03.004.

[52] Eldeeb MA, Akih-Kumgeh B. Investigation of 2,5-dimethyl furan and iso-octane ignition. Combust Flame 2015;162:2454–2465. doi:10.1016/j.combustflame.2015.02.013.

[53] Oehlschlaeger MA, Davidson DF, Herbon JT, Hanson RK. Shock Tube Measurements of Branched Alkane Ignition Times and OH Concentration Time Histories. Int J Chem Kinet 2004;36:67–78. doi:10.1002/kin.10173.

[54] Vermeer DJ, Meyer JW, Oppenheim AK. Auto-ignition of hydrocarbons behind reflected shock waves. Combust Flame 1972;18:327–36. doi:10.1016/S0010-2180(72)80183-4.

[55] Sakai Y, Ozawa H, Ogura T, Miyoshi A, Koshi M, Pitz WJ. Effects of Toluene Addition to Primary Reference Fuel at High Temperature. SAE Tech. Pap. Ser., 2010. doi:10.4271/2007-01-4104.

[56] Atef N, Kukkadapu G, Mohamed SY, Rashidi M Al, Banyon C, Mehl M, et al. A comprehensive iso-octane combustion model with improved thermochemistry and chemical kinetics. Combust Flame 2017;178:111–134. doi:10.1016/j.combustflame.2016.12.029.

[57] Dagaut P, Reuillon M, Cathonnet M. High pressure oxidation of liquid fuels from low to high temperature. 1. n-heptane and iso-octane. Combust Sci Technol 1993;95:233–60. doi:10.1080/00102209408935336.

[58] Curran HJ, Gaffuri P, Pitz WJ, Westbrook CK. A comprehensive modeling study of iso-octane oxidation. Combust Flame 2002;129:253–80. doi:10.1016/S0010-2180(01)00373-X.

[59] Law CK. Combustion physics. Cambridge: Cambridge University Press; 2006. doi:10.1017/CBO9780511754517.

[60] Bhattacharya A, Banerjee DK, Mamaikin D, Datta A, Wensing M. Effects of Exhaust Gas Dilution on the Laminar Burning Velocity of Real-World Gasoline Fuel Flame in Air. Energy and Fuels 2015;29:6768–79. doi:10.1021/acs.energyfuels.5b01299.

[61] Dirrenberger P, Glaude PA, Bounaceur R, Gall H Le, Pires A, Konnov AA, et al. Laminar burning velocity of gasolines with addition of ethanol. Fuel 2014;115:162–9. doi:10.1016/j.fuel.2013.07.015.

[62] Sileghem L, Alekseev VA, Vancoillie J, Nilsson EJK, Verhelst S, Konnov AA. Laminar burning velocities of primary reference fuels and simple alcohols. Fuel 2014;115:32–40. doi:10.1016/j.fuel.2013.07.004.

[63] Van Lipzig JPJ, Nilsson EJK, De Goey LPH, Konnov AA. Laminar burning velocities of n-heptane, iso-octane, ethanol and their binary and tertiary mixtures. Fuel 2011;90:2773–2781. doi:10.1016/j.fuel.2011.04.029.

[64] Kumar K, Freeh JE, Sung CJ, Huang Y. Laminar Flame Speeds of Preheated iso-

Octane/O2/N2 and n-Heptane/O2/N2 Mixtures. J Propuls Power 2007;23:428–436. doi:10.2514/1.24391.

[65] Wang X, Fan X, Yang K, Wang J, Jiao X, Guo Z. Laminar flame characteristics and chemical kinetics of 2-methyltetrahydrofuran and the effect of blending with isooctane. Combust Flame 2018;191:213–25. doi:10.1016/j.combustflame.2017.12.028.

[66] Konnov AA, Mohammad A, Kishore VR, Kim N Il, Prathap C, Kumar S. A comprehensive review of measurements and data analysis of laminar burning velocities for various fuel+air mixtures. Prog Energy Combust Sci 2018;68:197–267. doi:10.1016/j.pecs.2018.05.003.

[67] Egolfopoulos FN, Hansen N, Ju Y, Kohse-Höinghaus K, Law CK, Qi F. Advances and challenges in laminar flame experiments and implications for combustion chemistry. Prog Energy Combust Sci 2014;43:36–67. doi:10.1016/j.pecs.2014.04.004 Review.

[68] Zhang Z, Cheng P, Tan J, Liang J, Li Y, Li G. The uncertainty of laminar burning velocity of premixed H2-air flame induced by the non-uniform initial temperature field inside the constant-volume combustion vessel. Int J Hydrogen Energy 2018;43:21049–59. doi:10.1016/j.ijhydene.2018.09.037.

[69] Chen Z. On the accuracy of laminar flame speeds measured from outwardly propagating spherical flames: Methane/air at normal temperature and pressure. Combust Flame 2015;162:2442–2453. doi:10.1016/j.combustflame.2015.02.012.

[70] Ma X, Jiang C, Xu H, Ding H, Shuai S. Laminar burning characteristics of 2-methylfuran and isooctane blend fuels. Fuel 2014;116:281–291. doi:10.1016/j.fuel.2013.08.018.

[71] Andrae J, Johansson D, Björnbom P, Risberg P, Kalghatgi G. Co-oxidation in the auto-ignition of primary reference fuels and n-heptane/toluene blends. Combust Flame 2005;140:267–86. doi:10.1016/j.combustflame.2004.11.009.

[72] Andrae JCG, Björnbom P, Cracknell RF, Kalghatgi GT. Autoignition of toluene reference fuels at high pressures modeled with detailed chemical kinetics. Combust Flame 2007;149:2–24. doi:10.1016/j.combustflame.2006.12.014.

[73] Tripathi R, Burke U, Ramalingam AK, Lee C, Davis AC, Cai L, et al. Oxidation of 2-methylfuran and 2-methylfuran/n-heptane blends: An experimental and modeling study. Combust Flame 2018;196:54–70. doi:10.1016/j.combustflame.2018.05.032.

[74] Snitsiriwat S, Bozzelli JW. Thermochemical properties for isooctane and carbon radicals: Computational study. J Phys Chem A 2013;117:421−429. doi:10.1021/jp3041154.

[75] He X, Walton SM, Zigler BT, Wooldridge MS, Atreya A. Experimental investigation of the intermediates of isooctane during ignition. Int J Chem Kinet 2007;39:498–517. doi:10.1002/kin.20254.

[76] Bhattacharya A, Datta A, Wensing M. Laminar burning velocity and ignition delay time for premixed isooctane–air flames with syngas addition. Combust Theory Model 2017;21:228–47. doi:10.1080/13647830.2016.1215533.

[77] Bhattacharya A, Datta A. Laminar burning velocity of biomass-derived fuels and its significance in combustion devices. 2018. doi:10.1007/978-981-10-8393-8_16.

[78] Valera-Medina A, Xiao H, Owen-Jones M, David WIF, Bowen PJ. Ammonia for power. Prog Energy Combust Sci 2018;69:63–102. doi:10.1016/j.pecs.2018.07.001.

[79] Frenklach M, Wang H. Detailed modeling of soot particle nucleation and growth. Symp Combust 1991;23:1559–66. doi:10.1016/S0082-0784(06)80426-1.

[80] Frenklach M. Reaction mechanism of soot formation in flames. Phys Chem Chem Phys 2002;4:2028–37. doi:10.1039/b110045a.

[81] Matheu DM, Dean AM, Grenda JM, Green WH. Mechanism generation with integrated pressure dependence: A new model for methane pyrolysis. J Phys Chem A 2003;107:8552–65. doi:10.1021/jp0345957.

[82] Bhattacharya A, Basu S. An investigation into the heat release and emissions from counterflow diffusion flames of methane/dimethyl ether/hydrogen blends in air. Int J Hydrogen Energy 2019. doi:10.1016/j.ijhydene.2019.06.190.

[83] Zeldovich Y. The oxidation of nitrogen in combustion explosions. Acta Physicochim USSR 1946;21:577–628.

[84] Glarborg P, Miller JA, Ruscic B, Klippenstein SJ. Modeling nitrogen chemistry in combustion. Prog Energy Combust Sci 2018;67:31–68. doi:10.1016/j.pecs.2018.01.002.

[85] Bartok W, Engleman V, Goldstein R, del Valle E. Basic kinetic studies and modeling of nitrogen oxide formation in combustion processes. AIChE Symp Ser 1972;68:30–38.